\documentclass[12pt]{article}
\usepackage{amsfonts}
\usepackage{amssymb}
\usepackage{amsmath}
\usepackage{graphicx}
\usepackage{graphics}
\usepackage{color}
\usepackage{epsfig}
\newcommand{\be}{\begin{equation}}
\newcommand{\ee}{\end{equation}}
\newcommand{\bea}{\begin{eqnarray}}
\newcommand{\eea}{\end{eqnarray}}
\newcommand{\bwt}{\begin{widetext}}
\newcommand{\ewt}{\end{widetext}}

\renewcommand{\theequation}{\thesection.\arabic{equation}}
\newcounter{newapp}
\renewcommand{\thenewapp}{\Alph{newapp}}
\begin{document}
\begin{center}
{\large \bf AdS-Carroll Branes}
\end{center}
\begin{center}
{\bf T.E. Clark}\footnote{e-mail address: clarkt@purdue.edu}\\
{Department of Physics \& Astronomy,\\
 Purdue University,\\
 West Lafayette, IN 47907-2036, U.S.A.}\\
~\\
and \\
~\\
{\bf T. ter Veldhuis}\footnote{e-mail address: terveldhuis@macalester.edu}\\
{Van Swinderen Institute for Particle Physics and Gravity,\\
University of Groningen,\\
Nijenborgh 4, 9747 AG Groningen, The Netherlands \\
and \\
Department of Physics \& Astronomy,\\
 Macalester College,\\
 Saint Paul, MN 55105-1899, U.S.A.}
\end{center}
\vspace{0.5in.}
\begin{center}
{\bf Abstract}
\end{center}
\begin{flushleft}
Coset methods are used to determine the action of a co-dimension one brane (domain wall) embedded in (d+1)-dimensional AdS space in the Carroll limit in which the speed of light goes to zero.  The action is invariant under the non-linearly realized symmetries of the AdS-Carroll spacetime.  The Nambu-Goldstone field exhibits a static spatial distribution for the brane with a time varying momentum density related to the brane's spatial shape as well as the AdS-C geometry.  The AdS-C vector field dual theory is obtained.
\end{flushleft}
\pagebreak

\section{Introduction}

The symmetries of spacetime delimit the form of the action for fields on it.  The familiar case of Poincar\'e symmetric spacetime results in particle motion being restricted to the local forward lightcone.  This lightcone opens up to be the forward time half space in the Galilean limit in which the speed of light $c\rightarrow \infty$ and instantaneous interaction is possible.  On the other hand, as the speed of light vanishes, $c \rightarrow 0$, the causal lightcone closes to be just the forward time half line.  Such a contraction of spacetime is known as Carroll spacetime with symmetries generated by the Wigner-In\"on\"u contracted Poincar\'e algebra, $c\rightarrow 0$, to the Carroll algebra \cite{Levy-Leblond}, \cite{Bacry:1968zf}.  A particle in such a spacetime must remain stationary as the time axis is the lightcone.  This lack of motion can be found by considering the $c\rightarrow 0$ limit of its Poincar\'e geodesic action.  For a free particle moving in 1+1 dimensional Minkowski spacetime its action is given by
\bea
\Gamma &=& -mc^2 \int d\tau = -mc^2\int \sqrt{dt^2 - d{x}^2 /c^2} \cr
 &=& -mc^2 \int dt \sqrt{1-\dot{x}(t)^2/c^2} . 
\eea
Introducing a Lagrange multiplier auxiliary velocity $v(t)$
\be
\dot{x}(t)/c = \tanh{v(t)} ,
\ee
the action becomes
\be
\Gamma = -mc^2 \int dt \cosh{v(t)}\left[ 1 -\frac{1}{c}\dot{x}(t) \tanh{v(t)}\right] .
\ee
In order to take the Carroll limit $c\rightarrow 0$ the velocity is scaled by the speed of light $v(t)= 2c w(t)$ yielding the action
\be
\Gamma =-mc^2 \int dt \cosh{2cw(t)} \left[ 1 -\frac{1}{c}\dot{x}(t) \tanh{2cw(t)}\right].
\ee
Letting $c\rightarrow 0$ the Carroll limit for the action $\Gamma_C = \Gamma /mc^2 $ is obtained
\be
\Gamma_C = -\int dt \left[ 1- 2w(t) \dot{x}(t) \right].
\ee
As expected there is no causal relation between different events along the particle's trajectory and it remains stationary $\dot{x}(t) =0= \dot{w}(t)$ \cite{Bergshoeff:2014jla}, \cite{Bergshoeff:2015wma}.

Extending the limiting procedure to membranes inserted into (d+1)-dimensional Minkowski spacetime, the $c \rightarrow 0$ contraction yields the Carrollian Nambu-Goto action for the brane.  Such a limit occurs in the case of effective field theory of tachyon brane condensation in which the tachyon field rolls to the Carrollian limit \cite{Gibbons:2002tv}.  The Carroll brane action can be obtained from the contraction of the Nambu-Goto action for a one-codimensional brane
\bea
\Gamma_{\rm NG} &=& -\sigma \int d^d x \sqrt{-(-1)^{d}\det{g}}=-\sigma \int d^d x \sqrt{-(-1)^{d}\det{(\eta_{\mu\nu} - \partial_\mu \phi \partial_\nu \phi)}} \cr
 &=&-\sigma \int d^d x \sqrt{1 -\partial_\mu \phi \partial^\mu \phi} ,
\label{NGAction}
\eea
where the (d+1)-dimensional spacetime has been spontaneously broken to that of a d-dimensional world volume by the formation of a domain wall in the additional dimension.  These Poincar\'e symmetries are compactly described by the invariant interval $ds^2 = dx^\mu \eta_{\mu\nu} dx^\nu - dz^2$ with $z$ denoting the one-dimensional covolume coordinate and $x^\mu$ with $\mu = 0,1,\ldots , p=(d-1)$ denoting the $d$-dimensional world volume coordinates.  Replacing $z=\phi (x)$ it is obtained that $ds^2 =dx^\mu g_{\mu\nu}dx^\nu = dx^\mu \left(\eta_{\mu\nu} - \partial_\mu \phi \partial_\nu  \phi  \right)dx^\nu$ resulting in the $(d+1)$ dimensional space-time invariance of the Nambu-Goto action (\ref{NGAction}) with brane tension $\sigma$.  

In order to implement the Carrollian contraction, Lagrange multiplier auxiliary fields $V_\mu (x)$ are introduced so that
\be
\partial^\mu \phi = - \eta^{\mu\nu} V_\nu \frac{\tanh{\sqrt{V^2}}}{\sqrt{V^2}} ,
\label{multiplier}
\ee
with $V^2 = V_\mu \eta^{\mu\nu} V_\nu$ and $x_\mu = (ct , x_m)$ while $\partial^\mu = \frac{\partial}{\partial x_\mu}$.  The action becomes
\be
\Gamma_{\rm NG} = -c\sigma \int dt d^p x \cosh{\sqrt{V^2}}\left[ 1+ \left( V_\mu \frac{\tanh{\sqrt{V^2}}}{\sqrt{V^2}} \right)\partial^\mu \phi  \right] .
\label{NGGamma}
\ee
Making the speed of light explicit $V_0 =2c w$ and $V_m = 2 v_m$ for $m =1,2,\ldots , p= (d-1)$, the Carrollian limit $c\rightarrow 0$ of the action $\Gamma_{C} = -(1/c\sigma) \Gamma_{\rm NG}= \int dt d^p x {\cal L}_C$ is obtained
\be
\Gamma_{C} = \int dt d^p x {\cal L}_C = \int dt d^p x \cos{\sqrt{4 v^2}} \left[ 1+ \frac{\tan{\sqrt{4 v^2}}}{\sqrt{4 v^2}} \left( 2 w \dot{\phi} + 2 v_m \partial^m \phi  \right)   \right] ,
\label{CMinkowski}
\ee
where, after separation of space and time coordinates, the spatial metric is just $\delta^{mn}$. Only subscripts $x_m$, $v_m$ and superscript derivatives $\partial^m = \frac{\partial}{\partial x_m}$ will be used, with $\dot{\phi}= \frac{\partial}{\partial t} \phi$. In the case of a Goldstone field, as is $\phi$, the leading term in the derivative expansion of the action (thin wall limit) is uniquely fixed.  The field equations are found to be 
\bea
0 &=& \frac{\delta \Gamma_C}{\delta w(x, t)} = 2\dot{\phi} \left[\frac{\sin{\sqrt{4 v^2}}}{\sqrt{4v^2}}  \right]  \cr
0 &=& \frac{\delta \Gamma_C}{\delta v_m (x, t)} = 2 \cos{\sqrt{4v^2}} \left(\left[ 2 \delta^{nr} v_r \frac{\tan{\sqrt{4 v^2}}}{\sqrt{4 v^2}} -\partial^n \phi  \right] \times \right. \cr
 & &\left. \qquad\qquad\times\left[ \frac{\tan{\sqrt{4 v^2}}}{\sqrt{4 v^2}} P_{Tns} (v) + P_{Lns} (v)  \right] \delta^{sm} \right) \cr
  & &\qquad\qquad\qquad\qquad + 2\dot{\phi} \frac{w v_s \delta^{sm}}{v^2} \cos{\sqrt{4 v^2}} \left( \frac{\tan{\sqrt{4 v^2}}}{\sqrt{4 v^2}} -1  \right) \cr
0 &=& \frac{\delta \Gamma_C}{\delta \phi (x, t)} = -\frac{\partial}{\partial t} \left[2w \frac{\sin{\sqrt{4 v^2}}}{\sqrt{4 v^2}}  \right] - \frac{\partial}{\partial x_m} \left[2 v_m \frac{\sin{\sqrt{4 v^2}}}{\sqrt{4 v^2}}  \right] ,
\label{fieldeqs}
\eea
where the transverse $P_{Tmn} (v)$ and longitudinal $P_{Lmn} (v)$ projection matrices are defined as
\bea
P_{Tmn} (v) &=& \delta_{mn} - \frac{v_m v_n}{v^2}  \cr
P_{Lmn} (v) &=& \frac{v_m v_n}{v^2}  .
\eea
Although informally obtained in the introduction, these results also follow from the coset method of section 2 for the brane embedded in $AdS-C$ space when the flat (Minkowski) space-time limit $m^2 \to 0$ is taken and $AdS-C \stackrel{m^2 \to 0}{\longrightarrow} C$, Carroll spacetime.

The field equations reflect the Carroll spacetime symmetries yielding $\dot{\phi}=0$ from the $w$ equation of motion, $\delta \Gamma_C /\delta w =0$ so that the brane's initial spatial shape does not evolve as expected from the collapse of the light cone in this limit.  Having set $\dot{\phi}=0$, the spatial components of $v_m$ obey the constraint  
$2v_m \frac{\tan{\sqrt{4v^2}}}{\sqrt{4v^2}}=\partial^m \phi$ as dictated by the $v_m$ field equation, 
$\delta \Gamma_C /\delta v_m =0$.  Both field equations are consistent with the initial auxiliary velocity field equation (\ref{multiplier}) for $c\rightarrow 0$.  Finally the broken space translation symmetry in the (p+1) direction yields the time variation of the momentum density as given by the $\phi$ field equation $\delta \Gamma_C /\delta \phi =0$.  The momentum density $\Pi$ is defined by
\be
\Pi \equiv \partial {\cal L}_C/ \partial \dot{\phi} = 2w \frac{\sin{\sqrt{4 v^2}}}{\sqrt{4v^2}}  .
\ee
while the derivatives of ${\cal L}_C$ with respect to the spatial derivatives of $\phi$ are denoted
\be
\Pi_m = \partial {\cal L}_C/\partial\partial^m \phi = 2 v_m \frac{\sin{\sqrt{4 v^2}}}{\sqrt{4v^2}} .
\ee
Thus the $\phi$-field equation has the form of a current conservation equation. Indeed, the field equation is the spontaneously broken translation current conservation equation for the Carroll spacetime.  The corresponding Noether current has the conserved form as above
\be
\frac{\partial}{\partial t} \Pi + \frac{\partial}{\partial x_m} \Pi_m =0 .
\ee

The action, equation (\ref{CMinkowski}), is invariant under the Carroll transformations, obtained by contracting the Poincar\'e transformations as $c\to 0$, of the (d+1)-dimensional Carroll spacetime which include the unbroken d-dimensional worldvolume time and space translations, space rotations and boosts with respective parameters $\epsilon ,~a_m ,~\alpha_{mn} ,~\beta_m$ and additionally, now non-linearly realized, the broken space translation of the covolume which is just a $\phi$ shift symmetry, boosts in that direction and rotations in a worldvolume-covolume plane with respective parameters $\zeta ,~\lambda ,~\kappa_m$.  Exploiting the invariance of the Minkowski interval $ds^2 = dx^\mu \eta_{\mu\nu} dx^\nu - dz^2$ under $(d+1)$-dimensional Poincar\'e transformations
\be
x^\prime_M = x_M +\lambda_{MN} \eta^{NP}x_P + a_M
\ee
where $x_M = (ct, x_m , z)$ with $z=\phi (x)$ and $\lambda_{MN}=-\lambda_{NM}$, the coordinate and field Poincar\'e transformations have the form
\bea
t^\prime &=& t -\frac{1}{c}\lambda_{0n} x_n -\frac{1}{c}\lambda_{0z} \phi +\frac{1}{c}a_0  \cr
x_m^\prime &=& x_m +c\lambda_{0m} t -\lambda_{mn} x_n - \lambda_{mz} \phi + a_m \cr
\phi^\prime (x^\prime , t^\prime) &=& \phi (x ,t) + c\lambda_{z0} t -\lambda_{zm} x_m + a_z .
\eea
Contracting the Poincar\'e symmetry transformations to those of the Carroll symmetries requires a rescaling of the time components of the transformation parameters so that
\be
a_0 = c\epsilon \qquad ; \qquad \lambda_{0m} = c \beta_m \qquad ; \qquad \lambda_{0z} = c 2\lambda 
\ee
while the purely spatial components are unchanged and are denoted as
\be
\zeta = a_z \qquad ; \qquad \alpha_{mn} =-\lambda_{mn} \qquad ; \qquad 2\kappa_m =\lambda_{mz} .
\ee
The new parameters denote the Carroll transformation parameters.  The $c\to 0$ Carroll spacetime transformations of the coordinates and field $\phi$ are thus obtained and have the non-linear form (see Appendix A for the coset method derivation and the $AdS-C$ to Carroll spacetime $C$ limit to obtain equations (\ref{txphitransformations}) and (\ref{wvtransformations}))
\bea
t^\prime &=& t +\epsilon -2\lambda \phi - \beta_m x_m  \cr
x_m^\prime &=& x_m + a_m + \alpha_{mn} x_n -2 \phi \kappa_m \cr
\phi^\prime (x^\prime , t^\prime ) &=& \phi (x ,t) +\zeta + 2\kappa_m x_m .
\label{txphitransformations}
\eea
Applying these transformations to the auxiliary Lagrange multiplier field definition, equation (\ref{multiplier}), so that
\be
\partial_\mu^\prime \phi^\prime (x^\prime , t^\prime ) = - V_\mu^\prime (x^\prime , t^\prime )  \frac{\tanh{\sqrt{V^{\prime~2}}}}{\sqrt{V^{\prime~2}}} ,
\ee
with $V_\mu^\prime (x^\prime , t^\prime )= V_\mu ( x, t) +\Delta V_\mu (x ,t)$ where $\Delta V_\mu = (2c\Delta w , 2 \Delta v_m )$ yields the $w$ and $v_m$ auxiliary fields' Carroll transformations
\bea
w^\prime (x^\prime , t^\prime ) &=& w (x ,t)\left[1+ \frac{v_m \kappa_m}{v^2}\left( 1-\sqrt{4v^2}\cot{\sqrt{4v^2}}  \right)\right] \cr
 & &\qquad\qquad -\beta_m v_m + \lambda \sqrt{4v^2}\cot{\sqrt{4v^2}}  \cr
 & & \cr
v_m^\prime (x^\prime , t^\prime ) &=& v_m (x ,t) +\alpha_{mn} v_n +\left(\sqrt{4v^2}\cot{\sqrt{4v^2}}P_{Tmn}(v) + P_{Lmn}(v)   \right) \kappa_n  .\cr
 & & 
\label{wvtransformations}
\eea

Since the time and space transformations involve functions thereof, the differential form of equation (\ref{txphitransformations}) yields the general coordinate transformation $G = \partial (t^\prime , x^\prime )/\partial (t, x)$.  That is, recombining $t$ and $x_m$ in the matrix $X_M =(t , x_m)$ where now $M, N = 0, 1, \ldots ,p$, the transformations are given by
\bea
dX_M^\prime &=& (dt^\prime , dx_m^\prime ) \cr
 &=& dX_N G^N_{~M} = (dt G^0_{~0} + dx_n G^n_{~0} , dt G^0_{~m} + dx_n G^n_{~m})  ,
 \eea
 with
 \bea
 G^N_{~M} &=& \begin{pmatrix}
 	\frac{\partial t^\prime}{\partial t} & \frac{\partial x_m^\prime}{\partial t} \cr
 	\frac{\partial t^\prime}{\partial x_n} & \frac{\partial x_m^\prime}{\partial x_n}
 \end{pmatrix}^N_{~~M} \cr
  &=&\begin{pmatrix}
1-2\lambda \dot{\phi} & -2 \dot{\phi} \kappa_m \cr
(-\beta_n-2\lambda \partial^n \phi)~~ & (\delta_{nm} +\alpha_{nm}-2\partial^n \phi \kappa_m ) 
\end{pmatrix}^N_{~~M} .
\eea
The spacetime transformation Jacobian is $dt^\prime d^p x^\prime = dt d^p x \det{G}$.  On the other hand the action $\Gamma_C$ is invariant under the Carrollian symmetry transformations, thus
\bea
\Gamma_C^\prime &=& \int dt^\prime d^p x^\prime {\cal L}^\prime_C  (x^\prime , t^\prime ) = \int dt d^p x \det{G} {\cal L}^\prime_C  (x^\prime , t^\prime )\cr
 &=& \int dt d^p x {\cal L}_C  (x, t) = \Gamma_C  ,
\eea
so that 
\be
{\cal L}^\prime_C  (x^\prime , t^\prime ) = \det{G^{-1}} {\cal L}_C  (x, t) .
\ee

For these Carrollian transformations the Noether currents take the cuplet form of time and spatial component currents
\be
J_M =
\begin{pmatrix}
\Pi \delta \phi + \delta t {\cal L}_C \cr
\Pi_m \delta \phi + \delta x_m {\cal L}_C 
\end{pmatrix}_{M=(0, m)}
\ee
with $\delta t =t^\prime -t$ and $\delta x_m = x_m^\prime -x_m$ where the intrinsic transformation $\delta \phi$ is defined as
\be
\delta \phi \equiv \phi^\prime (x ,t) - \phi (x, t)
 =\Delta \phi (x, t) -\delta t \dot{\phi} - \delta x_m \partial^m \phi .
\ee
with the total variation given by 
\be
\Delta \phi (x, t) \equiv \phi^\prime (x^\prime ,t^\prime ) -\phi (x, t) .
\ee
Thus Noether's theorem is ($\varphi_i = \{\phi , w, v_m \}$)
\bea
\Delta {\cal L}_C &=& {\cal L}_C^\prime (x^\prime ,t^\prime ) -{\cal L}_C (x, t) \cr
 &=&\frac{\partial}{\partial t} J_0 + \frac{\partial}{\partial x_m} J_m -{\cal L}_C \frac{\partial \delta t}{\partial t} -{\cal L}_C \frac{\partial \delta x_m}{\partial x_m} + \frac{\delta \Gamma_C}{\delta \varphi_i}\delta\varphi_i 
\label{NoethersThm}
\eea
where the last term vanishes by the field equations $\delta \Gamma_C /\delta \varphi_i =0$.
The conserved currents (before use of the field equation constraints) are given by the pairs displayed in Table 1.  
The action is invariant as reflected by the vanishing or explicit cancellation of the $\Delta {\cal L}_C +{\cal L}_C \left(\frac{\partial \delta t}{\partial t} + \frac{\partial \delta x_m}{\partial x_m}\right)=0$ terms in Noether's theorem for each of the Carrollian symmetry transformations.  Hence $\partial^M J_M = \frac{\partial}{\partial t} J_0 + \frac{\partial}{\partial x_m} J_m = -\frac{\delta \Gamma_C}{\delta \varphi_i} \delta \varphi_i$.  The associated conserved charges are given by $Q= \int d^p x J_0$ where $\dot{Q} = - \int_{{S}\rightarrow \infty} \vec{J} \cdot d\vec{S} \longrightarrow 0$.  Once again the time evolution of the momentum density is contained in the broken translation current conservation equation and the invariance of the Lagrangian
\be
0= \dot{\Pi} + \partial^m \Pi_m = - \frac{\delta\Gamma_C }{\delta \phi} .
\ee
~\\
\begin{center}
\begin{tabular}{|c|c|c|}
\hline\hline
Transformation & Parameter & Noether Currents \\ \hline\hline
Time Translations & $\epsilon$ & ${\cal H} = \Pi \dot{\phi} -{\cal L}_C$ \\
     &    &$\qquad=\cos{\sqrt{4v^2}}+\Pi_m \partial^m \phi$  \\
     &    & $h_m = \Pi_m \dot{\phi}$ \\ \hline
Space Translations & $a_n$ & $T^n = \Pi \partial^n \phi$  \\
     &    & $T_{m}^{~n} = \frac{\partial{\cal L}_C}{\partial\partial^m \phi} \partial^n \phi - \delta_{m}^{~n} {\cal L}_C$\\
     &    & $\qquad=\delta_{m}^{~n}[\cos{\sqrt{4v^2}}+\Pi \dot{\phi} +\Pi_r \partial^r \phi]-\Pi_m \partial^n \phi$\\ \hline
Broken Space Translations & $\zeta$ & $z=\Pi$ \\
    &    &  $z_m =\Pi_m$ \\ \hline
Broken Boosts  & $\lambda$  & $l=\phi {\cal H}$ \\
    &      & $l_m = \phi h_m$  \\ \hline
Broken Rotations & $\kappa_n$  & $K^n = 2\phi T^n + 2\delta^{nr} x_r \Pi$ \\
    &      & $K_{m}^{~n} = 2\phi T_{m}^{~n} +2 \Pi_m \delta^{nr} x_r$\\ \hline
Unbroken Rotations & $\alpha_{rs}$ & ${\cal M}^{rs} = x_r T^s - x_s T^r$ \\
    &      &  ${\cal M}_{m}^{~rs} = x_r T_{m}^{~s} - x_s T_{m}^{~r}$\\ \hline
Unbroken Boosts & $\beta_n$  & $B^n=  \delta^{nr} x_r {\cal H}$\\
      &     & $B^{~n}_{m} = \delta^{nr} x_r h_m$\\ \hline\hline
\end{tabular}
\end{center}
\begin{center}
{\bf Table 1} Carroll Spacetime Transformations and Noether Currents.
\end{center}

In summary, the $w$-field equation of motion, equation (\ref{fieldeqs}), yields the frozen spatial distribution of the domain wall as expected from the collapse of the lightcone to the positive time half line in the Carrollian $c\rightarrow 0$ limit: $\frac{\partial}{\partial t} \phi ( x, t) =0$.  Along with this the Lagrange multiplier $v_m$-field equation of motion simply reproduces the constraint of the \lq\lq inverse Higgs mechanism" \cite{Ivanov:1975zq}
\be
2v_m \frac{\tan{\sqrt{4 v^2}}}{\sqrt{4 v^2}} =\partial^m \phi .
\ee
These field equation constraints also follow directly from the Lagrange multiplier equation (\ref{multiplier}) in the $c\rightarrow 0$ limit.  Finally, although the brane is stationary, the momentum must vary in time in order to balance the tension due to the domain wall's local spatial shape where using the Lagrange multiplier constraints so that $\dot{\phi} =0$ and $ 2v_m \frac{\tan{\sqrt{4 v^2}}}{\sqrt{4 v^2}} = \partial^m \phi$, it is found that
\be
\dot{\Pi} = -\partial^m \left[ \frac{\partial^m \phi}{\sqrt{1+ \partial^s \phi \partial^s \phi}} \right] .
\ee

The purpose of this paper is to determine the Carrollian limit for branes in $AdS$ spacetime \cite{Clark:2005ht}, \cite{Clark:2007rn}.  The $D=d+1$ dimensional $AdS$ spacetime symmetry algebra is contracted in the Carroll limit, $c\rightarrow 0$.  For a p-brane action in the alternate string Carrollian limit of Minkowski space see \cite{cardona}. Application of the Carrollian limit to gravity and electromagnetism is discussed in \cite{Hartong:2015xda} and \cite{Duval:2014uoa}. 

In Section 2 coset methods \cite{Coleman:1969sm}, \cite{Volkov73}, \cite{Ivanov:1975zq} are applied to the $AdS-C$ algebra for the case of an embedded co-dimension one p-brane (domain wall).  The induced vielbeine, covariant derivatives and spin connections are determined using the Maurer-Cartan one-form associated with the p-brane coset element.  The action is constructed and shown to be invariant under the non-linearly realized $AdS-C_{d+1}$ broken to $AdS-C_{d}$ symmetries by the brane embedding.  The symmetry transformations are detailed in Appendix A.  Alternatively, the $AdS-C$ action can be obtained by making the speed of light $c$ dependence explicit in the $AdS_{d+1}\rightarrow AdS_d$ case and taking the $c\to 0$ limit.  Using the results of reference \cite{Clark:2005ht} this approach is demonstrated in Appendix B. 

From the action the field equations are determined.  As expected due to the collapse of the forward light cone to the positive time half-line, the spatial shape of the brane is stationary.  However, the spatial shape of the brane as well as the $AdS-C$ geometry requires its conjugate momentum density to be time dependent.  Finally Noether's theorem is applied to the broken space translation symmetry in order to calculate the current and its conservation equation.  Section 3 presents the action in terms of a product of the background $AdS-C_d$ world volume vielbein and the Nambu-Goto-Carrollian vielbein.  This is then used to express the action in terms of its dual vector theory.  The results of the brane embedding are reviewed in the section 4 conclusions.
\pagebreak
\setcounter{newapp}{2}
\setcounter{equation}{0}

\section{AdS-Carroll Space And The Coset Method}

The AdS-Carroll spacetime is defined by the Wigner-In\"on\"u contraction of the AdS symmetry algebra for the speed of light vanishing, $c\rightarrow 0$.  The isometry group of the $D$-dimensional $AdS$ space is given by the $SO(2, D-1)$ algebra of symmetry generators with the commutation relations
\bea
\left[ M^{MN}, M^{RS} \right] &=& -i\left( \eta^{MR} M^{NS} -\eta^{MS} M^{NR} +\eta^{NS} M^{MR} - \eta^{NR} M^{MS}  \right) \cr
\left[ M^{MN}, P^L \right] &=& i\left( P^M \eta^{NL}-P^N \eta^{ML} \right) \cr
\left[ P^M , P^N \right] &=& -im^2 M^{MN} ,
\label{AdSD}
\eea
where $L, M, N, R, S = 0,1,2,\ldots , D-1$ and the metric $\eta_{MN} = (+1,-1,-1,\ldots , -1)$ with $m^2 = 1/R^2$ and $R$ the curvature of the AdS hyperboloid.

Introducing the explicit factors of the speed of light for the time related components, the time component involved generators $H$ and $B^A$ are defined as
\bea
P^0 &=& \frac{1}{c} H \cr
M^{A0} &=& \frac{1}{c} B^A ,
\eea
while the spatial components remain unscaled $P^A$, $M^{AB}$,
\bea
P^A&=& P^A \cr
M^{AB} &=& M^{AB},
\eea
for $A,~B,~C,~D = 1,2,\ldots ,(D-1)$ denoting the spatial indices.  The $SO(2, D-1)$ algebra contracts to the AdS-Carroll algebra in the $c\rightarrow 0$ limit
\bea
\left[M^{AB}, M^{CD}\right] &=& +i\left( \delta^{AC} M^{BD} -\delta^{AD} M^{BC}+\delta^{BD} M^{AC}-\delta^{BC} M^{AD} \right) \cr
\left[M^{AB}, B^C\right] &=& -i\left( B^A \delta^{BC} -B^B \delta^{AC} \right)  \cr
\left[M^{AB}, P^C\right] &=& -i\left( P^A \delta^{BC} -P^B \delta^{AC} \right)  \cr
\left[B^A, P^B \right] &=& +i \delta^{AB} H \cr
\left[H , P^A\right] &=& +im^2 B^A  \cr
\left[P^A , P^B \right] &=& -im^2 M^{AB}  ,
\eea
with remaining commutators vanishing.

A brane embedded in this AdS-Carroll spacetime will break its $AdS-C_D$ symmetries down to those of the $d$-dimensional worldvolume $AdS-C_d$ and its complementary covolume with the remaining symmetries being spontaneously broken.  In the case considered here, the insertion of a domain wall results in a $d = (1+p)= (D-1)$-dimensional worldvolume and 1-dimensional covolume.  Choosing the $(p+1)^{\rm th}$ spatial direction as the broken translation symmetry direction the AdS-Carroll algebra can be expressed in terms of broken and unbroken generators with the generators $H, P^m , M^{mn} , B^m $, with $m,n = 1,2, \ldots , p$, as unbroken generators and $P^{p+1} \equiv Z , M^{p+1,~m} \equiv \frac{1}{2} K^m , B^{p+1} \equiv \frac{1}{2}L$ as the broken generators.  $M^{mn}$ are the $SO(p)$ worldvolume spatial rotation generators while the worldvolume spatial translation generators $P^m$ form an $SO(p)$ vector with time translations generated by $H$.  The $SO(p)$ vector $B^m$ generates worldvolume boosts in the $m$-direction.  The translations in the covolume spatial direction are generated by $Z$ while boosts in that direction are generated by $L$.  Finally broken rotations in the covolume-$m$ worldvolume plane are generated by the $SO(p)$ vector $K^m$.  Consequently the $AdS-C_{D=d+1}$ algebra can be expressed in terms of these worldvolume and domain wall charges.  The $AdS-C_{d=p+1}$ worldvolume isometries are given by the $H, P^m ,M^{mn} , B^m $ algebra (only nontrivial commutators listed)
\bea
\left[M^{mn}, M^{rs}\right] &=& +i\left( \delta^{mr} M^{ns} -\delta^{ms} M^{nr}+\delta^{ns} M^{mr}-\delta^{nr} M^{ms} \right) \cr
\left[M^{mn}, B^l\right] &=& -i\left( B^m \delta^{nl} -B^n \delta^{ml} \right)  \cr
\left[M^{mn}, P^l\right] &=& -i\left( P^m \delta^{nl} -P^n \delta^{ml} \right)  \cr
\left[B^m, P^n \right] &=& +i \delta^{mn} H \cr
\left[H , P^m\right] &=& +im^2 B^m  \cr
\left[P^m , P^n \right] &=& -im^2 M^{mn}  .
\label{algebra1}
\eea
The broken symmetry generators $Z, L, K^m$ commute with the unbroken generators above according to their unbroken subgroup representation
\bea
\left[M^{mn} , Z \right] &=& 0 \cr
\left[M^{mn} , L \right] &=& 0 \cr
\left[M^{mn}, K^l\right] &=& -i\left( K^m \delta^{nl} -K^n \delta^{ml} \right)  \cr
\left[B^{m} , Z \right] &=& 0 \cr
\left[B^{m} , L \right] &=& 0 \cr
\left[B^m , K^n\right] &=& +i \delta^{mn} L  \cr
\left[P^{m} , Z \right] &=& \frac{i}{2} m^2 K^m \cr
\left[P^{m} , L \right] &=& 0 \cr
\left[P^m , K^n\right] &=& +2i \delta^{mn} Z  \cr
\left[H , Z \right] &=& \frac{i}{2} m^2 L \cr
\left[H, L \right] &=& 0 \cr
\left[H , K^n\right] &=& 0 .
\label{algebra2}
\eea
Finally the broken charges $Z, L, K^m$ commute amongst themselves to yield the charges of the unbroken subalgebra 
\bea
\left[ Z, L \right] &=& -2i H  \cr
\left[ Z, K^m \right] &=& -2i P^m  \cr
\left[ L, K^m \right] &=& -4i B^m  \cr
\left[ K^m, K^n \right] &=& 4i M^{mn}  .
\label{algebra3}
\eea

The domain wall spontaneously breaks the $AdS-C_D$ spacetime symmetries down to those of the $AdS-C_d$ worldvolume.  As a derivative expansion the leading form of the brane action is uniquely determined.  The Goldstone boson fields $\phi( x, t)$ corresponding to the long wavelength oscillations of the domain wall parameterize the coset coordinates along with the fields associated with the broken boost and rotations, $w(x,t)$ and $v_m (x,t)$, respectively.  The geometry of the underlying $AdS-C_d$ worldvolume spacetime is described by the time $t$ and space $x_m$, $m= 1,2, \ldots, p=(d-1)$, coordinate group elements.  Overall these fields and spacetime coordinates parameterize the $AdS-C_D/ISO(p)$ coset element $\Omega$ (note generators are defined with superscripts)
\be
\Omega \equiv e^{itH+ix_m P^m} e^{i\phi (x,t) Z} e^{iw(x,t) L +i v_m (x,t) K^m} 
\label{AdSCcoset}
\ee
where ISO(p) is the unbroken subgroup with generators $M^{mn}$ and $B^m$. The background worldvolume coset $\bar{\Omega}\in AdS-C_d /ISO(p)$
\be
\bar{\Omega}\equiv e^{itH+ix_m P^m}
\label{backgroundcoset}
\ee
is used to determine the $AdS-C_d$ background vielbeine and spin connections via the Maurer-Cartan 1-form $\tilde{\bar{\omega}}$
\be
\tilde{\bar{\omega}} \equiv -i \bar{\Omega}^{-1} d\bar{\Omega}= \bar{\omega}_H H +\bar{\omega}_{Pa} P^a + \frac{1}{2}\bar{\omega}_{ab} M^{ab} + \bar{\omega}_{Ba} B^a .
\ee
Expanding the 1-forms in terms of the coordinate differentials the Maurer-Cartan 1-form becomes (with tangent space indices denoted $a, b= 1, 2, \ldots , p$ and world volume indices denoted $m, n= 1,2, \ldots ,p$)
\bea
\tilde{\bar{\omega}} &=& (dt \bar{e}^0_{~0} + dx_m \bar{e}^m_{~0}) H + (dt \bar{e}^0_{~a} + dx_n \bar{e}^n_{~a}) P^a \cr
 & &\qquad + \frac{1}{2} \left( dt \bar{\omega}_{ab}^t +dx_r \bar{\omega}_{ab}^r \right) M^{ab} + \left( dt \bar{\omega}_{0a}^t +dx_r \bar{\omega}_{0a}^r \right) B^a ,
\eea
where the background vielbeine are found to be
\bea
\bar{e}^{0}_{~0} &=& \frac{\sinh{\sqrt{m^2 x^2}}}{\sqrt{m^2 x^2}} \cr
\bar{e}^{m}_{~0} &=& \frac{x_m t}{x^2} \left(1- \frac{\sinh{\sqrt{m^2 x^2}}}{\sqrt{m^2 x^2}}  \right) \cr
\bar{e}^{n}_{~a} &=& \left( \frac{\sinh{\sqrt{m^2 x^2}}}{\sqrt{m^2 x^2}}  \right) P_{Tna} (x) + P_{Lna} (x) \cr
\bar{e}^{0}_{~a} &=& 0
\label{backgrdvielbeine}
\eea
with $x^2 = x_m \delta^{mn} x_n = x_m x_m $.  The background spin connections are also obtained as
\bea
\bar{\omega}_{ab}^t &=& 0 \cr
\bar{\omega}_{ab}^r &=& \left(\frac{1-\cosh{\sqrt{m^2 x^2}}}{x^2}\right) \left( \delta_{a}^{s}\delta_b^r - \delta_{b}^{s}\delta_a^r\right) x_s \cr
\bar{\omega}^t_{0a} &=& \left(\frac{1-\cosh{\sqrt{m^2 x^2}}}{x^2}\right) x_a \cr
\bar{\omega}^r_{0a} &=& -\left(\frac{1-\cosh{\sqrt{m^2 x^2}}}{x^2}\right) t \delta_{a}^{~r} .
\label{backgspinconnections}
\eea

The $AdS-C_d$ background vielbein $\bar{E}^M_{~A}$ is defined as the matrix relating the coordinate differentials $dX_N=(dt, dx_n)$, with $M, N= 0, 1, \ldots , p$, to the covariant coordinate differentials $\tilde{\bar{\omega}}_A=(\bar{\omega}_H, \bar{\omega}_{Pa})$, with $A, B= 0, 1, \ldots , p$ as well and $\tilde{\bar{\omega}}_0 =\bar{\omega}_H$ and $\tilde{\bar{\omega}}_a =\bar{\omega}_{Pa}$, thus
\bea
\tilde{\bar{\omega}}_A &=& \left(\bar{\omega}_H ~~~ \bar{\omega}_{Pa} \right) =dX_M \bar{E}^M_{~A} \cr
 &=&\left( dt ~~~ dx_m \right)
\begin{pmatrix}
\bar{e}^0_{~0} & {0} \cr
\bar{e}^m_{~0} & \bar{e}^m_{~a}\cr
\end{pmatrix}  ,
\label{backgroundmatrixE}
\eea
that is
\be
\bar{E}^M_{~A} = \begin{pmatrix}
\bar{e}^0_{~0} & \bar{e}^0_{~a}={0} \cr
\bar{e}^m_{~0} & \bar{e}^m_{~a}\cr
\end{pmatrix} 
= \begin{pmatrix}
\bar{E}^0_{~0} & \bar{E}^0_{~a}={0} \cr
\bar{E}^m_{~0} & \bar{E}^m_{~a}\cr
\end{pmatrix} ,
\ee
with $\det{\bar{E}} =\bar{e}^0_{~0} \det{\bar{e}^m_{~a}}$, where $a, b, m, n=1,2, \ldots , p$.

On the otherhand, the Maurer-Cartan 1-form for the domain wall breakdown of $AdS-C_D\rightarrow AdS-C_d$ can be constructed using the coset element $\Omega$
\bea
\tilde{\omega} &=& -i \Omega^{-1} d \Omega \cr
 &=& \omega_H H +\omega_{Pa} P^a +\omega_Z Z + \omega_L L + \omega_{Ka} K^a +\frac{1}{2} \omega_{Mab} M^{ab} + \omega_{Ba} B^a .\cr
 & & 
\eea
This yields the vielbeine, covariant derivatives of the fields and spin connections.  The vielbeine are given in terms of the coordinate differentials according to 
\bea
\omega_H &=& dt e^0_{~0} + dx_m e^m_{~0} \cr
\omega_{Pa} &=& dt e^0_{~a} + dx_m e^m_{~a} ,
\eea
with $v^2 = v_a \delta^{ab} v_b =v_a v_a$ and
\bea
e^0_{~0} &=& \bar{e}^0_{~0} \cosh{\sqrt{m^2 \phi^2}} + 2w \dot{\phi} \left( \frac{\sin{\sqrt{4v^2}}}{\sqrt{4v^2}}  \right)  \cr
e^m_{~0} &=& \bar{e}^m_{~0} \cosh{\sqrt{m^2 \phi^2}} + 2w \partial^m {\phi} \left( \frac{\sin{\sqrt{4v^2}}}{\sqrt{4v^2}}  \right) \cr
 & &\qquad\qquad +\cosh{\sqrt{m^2 \phi^2}}\left(\frac{\cos{\sqrt{4v^2}}-1}{{v^2}}  \right)wv_a \bar{e}^m_{~a} \cr
e^0_{~a} &=& 2 \dot{\phi} v_a \left( \frac{\sin{\sqrt{4v^2}}}{\sqrt{4v^2}}  \right)  \cr
e^m_{~a} &=& 2 \partial^m {\phi} v_a\left( \frac{\sin{\sqrt{4v^2}}}{\sqrt{4v^2}}  \right) \cr
 & &\qquad + \cosh{\sqrt{m^2 \phi^2}} \left[ P_{Tab} (v) + ( \cos{\sqrt{4v^2}} )P_{Lab} (v) \right]  \bar{e}^m_{~b} .\cr
 & & 
\label{vielbeine}
\eea
The $AdS-C_D$ spacetime vielbein ${E}^M_{~A}$ is defined as the matrix relating the coordinate differentials $dX_M=(dt, dx_m)$ to the covariant coordinate differentials $\tilde{\omega}_A = ({\omega}_H, {\omega}_{Pa})$, that is $\tilde{\omega}_0 = \omega_H$ and $\tilde{\omega}_a = \omega_{Pa}$, thus
\bea
\tilde{\omega}_A &=&\left({\omega}_H ~~~ {\omega}_{Pa} \right) = dX_M E^M_{~A} \cr
 &=&\left( dt ~~~ dx_m \right)
\begin{pmatrix}
{e}^{0}_{~0} & {e}^0_{~a} \cr
{e}^{m}_{~0} & {e}^{m}_{~a}\cr
\end{pmatrix} ,
\label{matrixE}
\eea
that is
\be
E^M_{~A} = 
\begin{pmatrix}
	{E}^{0}_{~0} & {E}^0_{~a} \cr
	{E}^{m}_{~0} & {E}^{m}_{~a}\cr
\end{pmatrix} = \begin{pmatrix}
{e}^{0}_{~0} & {e}^0_{~a} \cr
{e}^{m}_{~0} & {e}^{m}_{~a}\cr
\end{pmatrix} .
\label{matrixEe2}
\ee

The brane field's covariant derivatives, $\nabla^t \phi$ and $\nabla^m \phi$, are given by the $\omega_Z$ one-form
\be
\omega_Z = dt \nabla^t \phi + dx_m \nabla^m \phi  
\ee
where 
\bea
\nabla^t \phi &=& \dot{\phi} \cos{\sqrt{4v^2}} \cr
\nabla^m \phi &=& \bar{e}^m_{~a}\cosh{\sqrt{m^2 \phi^2}}\cos{\sqrt{4v^2}}\left(\frac{\bar{e}^{-1a}_{~~~n} \frac{\partial}{\partial x_n} \phi}{\cosh{\sqrt{m^2 \phi^2}}}  -2 \delta^{ab} v_b   \left( \frac{\tan{\sqrt{4v^2}}}{\sqrt{4v^2}}  \right)\right).\cr
 & & 
\label{covderivphi}
\eea
Likewise the auxiliary fields $w$ and $v_a$ have covariant derivatives determined by $\omega_L$ and $\omega_{Ka}$
\bea
\omega_L &=& dt \nabla^t w + dx_m \nabla^m w \cr
\omega_{Ka} &=& dt \nabla^t v_a + dx_m \nabla^m v_a 
\eea
where the derivatives are found to be
\bea
\nabla^t w &=& \dot{w} + \left(\frac{\sin{\sqrt{4v^2}}}{\sqrt{}4v^2} -1 \right) \left( \dot{w} -w \frac{v_a \dot{v}_a}{v^2}  \right) -\frac{\sin{\sqrt{4v^2}}}{\sqrt{4v^2}} \bar{\omega}_{0a}^t v_a \cr
 & & \qquad\qquad\qquad -\frac{1}{2}m^2 \phi \frac{\sinh{\sqrt{m^2 \phi^2}}}{\sqrt{m^2 \phi^2}} \cos{\sqrt{4v^2}} \bar{e}^0_{~0} \cr
\nabla^m w &=& \partial^m w + \left(\frac{\sin{\sqrt{4v^2}}}{\sqrt{}4v^2} -1 \right) \left( \partial^m {w} -w \frac{v_a \partial^m{v_a}}{v^2}  \right) -\frac{\sin{\sqrt{4v^2}}}{\sqrt{4v^2}} \bar{\omega}_{0a}^m v_a \cr
  & & \qquad -\frac{1}{2}m^2 \phi \frac{\sinh{\sqrt{m^2 \phi^2}}}{\sqrt{m^2 \phi^2}} \left[\cos{\sqrt{4v^2}} \bar{e}^m_{~0} +\frac{(1-\cos{\sqrt{4v^2}})}{v^2} w v_a \bar{e}^m_{~a}\right]\cr
\nabla^t v_a &=& \dot{v}_a +\left(\frac{\sin{\sqrt{4v^2}}}{\sqrt{}4v^2} -1 \right)P_{Tab} (v) \dot{v}_b +
\frac{\sin{\sqrt{4v^2}}}{\sqrt{4v^2}} \bar{\omega}_{ab}^t v_b  \cr
\nabla^m v_a &=& \partial^m {v}_a +\left(\frac{\sin{\sqrt{4v^2}}}{\sqrt{}4v^2} -1 \right)P_{Tab} (v) \partial^m {v}_b + \frac{\sin{\sqrt{4v^2}}}{\sqrt{4v^2}} \bar{\omega}_{ab}^s v_b \cr
 & & \qquad -\frac{1}{2}m^2 \phi \frac{\sinh{\sqrt{m^2 \phi^2}}}{\sqrt{m^2 \phi^2}} \left[ \cos{\sqrt{4v^2}} P_{Tab} (v) + P_{Lab} (v)  \right]  \bar{e}^m_{~b} .\cr
 & & 
\label{covderivwv}
\eea
Finally the spin connections are obtained from $\omega_M$ and $\omega_B$
\bea
\omega_{Mab} &=& dt \omega_{ab}^t + dx_m \omega_{ab}^m \cr
 &=& \bar{\omega}_{ab} + m^2 \phi \frac{\sinh{\sqrt{m^2 \phi^2}}}{\sqrt{m^2 \phi^2}} \frac{\sin{\sqrt{4v^2}}}{\sqrt{4v^2}}  \cr
 & &+ \left( 1- \cos{\sqrt{4v^2}}  \right) \left[P_{Lac} (v) \bar{\omega}_{bc} -P_{Lbc} (v) \bar{\omega}_{ac} - \left(\frac{dv_a v_b -dv_b v_a}{v^2}  \right)\right] \cr
\omega_{Ba} &=& dt \omega_{~a}^{t} + dx_m \omega_{~a}^{m} \cr
 &=& \bar{\omega}_{Ba} + \left( \frac{1- \cos{\sqrt{4v^2}}}{v^2}  \right)\left[ dw v_a -w dv_a -\bar{\omega}_{ab} w v_b - \bar{\omega}_{Bb} v_b v_a \right]  \cr
 & &+ 2 m^2 \phi \frac{\sinh{\sqrt{m^2 \phi^2}}}{\sqrt{m^2 \phi^2}} \frac{\sin{\sqrt{4v^2}}}{\sqrt{4v^2}}\left[w \bar{e}^m_{~a} dx_m - v_a \bar{e}^0_{~0} dt -v_a dx_m \bar{e}^m_{~0}   \right]  .\cr
 & & 
\eea
The $AdS-C_D$ invariant action is constructed in terms of the vielbein $E$
\be
\Gamma_{AdS-C_D} = \int dt d^p x {\cal L}_{AdS-C_D}(x, t) = \int dt d^p x \det{E} .
\ee

The $AdS-C_D$ transformations are non-linearly realized according to the group multiplication properties involving the coset $\Omega$ as detailed in Appendix A.  The invariance of the action follows from the transformation properties of the vielbein $E$.  The Maurer-Cartan one-forms transform according to which representation of the local $ISO(p)$ tangent space transformations that the associated operator belongs given by the unbroken subgroup element $h (x, t)$ obtained in the Appendix A.  Using the coset transformation law $g\Omega (x, t) = \Omega^\prime (x^\prime , t^\prime ) h( x, t)$, the one-forms transform as $\omega^\prime (x^\prime , t^\prime ) = h(x, t) \omega (x, t) h^{-1} (x, t) -i h(x, t) d h^{-1} (x, t)$, yielding
\bea
\omega^\prime &=& \omega_H^\prime H + \omega_{Pa}^\prime P^a +\omega_Z^\prime Z + \omega_L^\prime L +\omega_{Ka}^\prime K^a  + \frac{1}{2} \omega_{Mab}^\prime M^{ab} + \omega_{Ba}^\prime B^a \cr
 &=&h \omega h^{-1} -i h dh^{-1} \cr
 &=& \omega_H hHh^{-1} +\omega_{Pa}h P^a h^{-1}+\omega_Z hZ h^{-1}+ \omega_L hLh^{-1} +\omega_{Ka}h K^a h^{-1} \cr
 & &+ \frac{1}{2} \omega_{Mab} hM^{ab}h^{-1} + \omega_{Ba} hB^a h^{-1} -\frac{1}{2} d\theta_{ab} M^{ab} -d\theta_a B^a  . \cr
 & & 
\eea
Hence the one-forms' variations are obtained
\bea
\omega_H^\prime &=& \omega_H - \omega_{Pa}\theta_a \cr
\omega_{Pa}^\prime &=& \omega_{Pb} (\delta_{ba} - \theta_{ba} ) \equiv \omega_{Pb} R^{-1}_{ba} \cr
\omega^\prime_Z &=& \omega_Z \cr
\omega_L^\prime &=& \omega_L - \omega_{Ka} \theta_a \cr
\omega_{Ka}^\prime &=& \omega_{Kb} R^{-1}_{ba} \cr
\omega_{Mcd}^\prime &=& \omega_{Mab} R^{-1}_{ac} R^{-1}_{db} -d\theta_{cd} \cr
\omega_{Ba}^\prime &=& \omega_{Bb} R^{-1}_{ba} -d\theta_a +\frac{1}{2}\omega_{Mcd} (\theta_c \delta_{da} -\theta_d \delta_{ca} ) .
\label{oneformvariations}
\eea
The covariant coordinate differentials and vielbeine transform as
\bea
\tilde{\omega}_A^\prime &=& (\omega_H^\prime ~~\omega_{Pa}^\prime ) = (\omega_H ~~\omega_{Pb} )
\begin{pmatrix}
1&0 \cr
-\theta_b & R^{-1}_{ba} \cr
\end{pmatrix} \cr
 &=& dX_M^\prime E^{\prime M}_{~~A} = ( dt^\prime ~~ dx_m^\prime )
\begin{pmatrix}
e^{\prime 0}_{~~0} & e^{\prime 0}_{~~a}\cr
e^{\prime m}_{~~0} & e^{\prime m}_{~~a}
\end{pmatrix}
\cr
 &=& (dt ~~dx_r ) 
\begin{pmatrix}
e^0_{~0} & e^0_{~b} \cr
e^r_{~0}  & e^r_{~b} \cr
\end{pmatrix}
\begin{pmatrix}
1 & 0 \cr
-\theta_b & R^{-1}_{ba} \cr
\end{pmatrix} 
= dX_R E^R_{~B} \Lambda^B_{~A} ,
\label{HPomegamatrix}
\eea
with letters from the beginning of the alphabet denoting the tangent space transformation properties.  From Appendix A the coordinate differentials transform according to the general coordinate transformation
\be
dX^\prime_M = (dt^\prime ~~ dx_m^\prime ) = (dt ~~ dx_n ) 
\begin{pmatrix}
G^0_{~0} & G^0_{~m} \cr
G^n_{~0} & G^n_{~m} \cr
\end{pmatrix}
= dX_N G^N_{~M}
\label{Gtransformations}
\ee
where the complicated general coordinate transformation matrix is denoted by $G^N_{~M}$, with letters in the middle of the alphabet indicating world volume coordinate transformations.  The spacetime differentials have the Jacobian $dt^\prime d^p x^\prime = dt d^p x \det{G}$.  Thus the vielbein $E^M_{~A}$ transforms as
\be
E^{\prime M}_{~~A} = G^{-1M}_{~~~~N} E^N_{~B} \Lambda^B_{~A} 
\label{ETrans}
\ee
where the tangent space transformations have been denoted by
\be
\Lambda^B_{~A} = \begin{pmatrix}
1 & 0 \cr
-\theta_b & R^{-1}_{ba} \cr
\end{pmatrix} .
\label{Lambda}
\ee
Noting that $\det{\Lambda} = 1$ so that $\det{E^\prime} = \det{G^{-1}} \det{E}$, the action is invariant
\bea
\Gamma_{AdS-C_D}^\prime &=& \int dt^\prime d^p x^\prime \det{E^\prime} = \int dt d^p x \det{G} \det{G^{-1}} \det{E} \cr
 &=& \int dt d^p x \det{E} = \Gamma_{AdS-C_D} .
\eea

With the vielbeine in equation (\ref{vielbeine}) the $AdS-C_D$ invariant action is found
\be
\Gamma_{AdS-C_D} = \int dt d^p x \det{E} =\int dt d^p x \det{(e^m_{~a})} \left[ e^0_{~0} - e^0_{~a} e^{-1a}_{~~~~n} e^n_{~0} \right],
\ee
with (noting that $\bar{e}^{-1a}_{~~~~m} \bar{e}^m_{~0} = \delta^a_{~m} \bar{e}^m_{~0} = \bar{e}^a_{~0}$)
\bea
\det{E} &=& \det{\bar{E}} \cosh^p{\sqrt{m^2 \phi^2}} \cos{\sqrt{4v^2}} \left\{ \cosh{\sqrt{m^2 \phi^2}} + {\cal D}^a \phi 2 v_a \frac{\tan{\sqrt{4v^2}}}{\sqrt{4v^2}}   \right. \cr
 & &\left. \qquad\qquad\qquad\qquad +2 \bar{e}^{0~-1}_{~0} \dot{\phi} (w -v_a \bar{e}^a_{~0} ) \frac{\tan{\sqrt{4v^2}}}{\sqrt{4v^2}}  \right\} .
\label{ELagrangian}
\eea
The background $AdS-C_d$ spacetime measure is given by $\det{\bar{E}}= \bar{e}^0_{~0} \det{\bar{e}^m_{~a}}$ and the partially covariant spatial derivative ${\cal D}^a \phi = \bar{e}^{-1a}_{~~~m} \frac{\partial}{\partial x_m}\phi$.
Note that the covariant derivatives of $\phi$, equation (\ref{covderivphi}), are given by
\bea
\nabla^t \phi &=& \dot{\phi} \cos{\sqrt{4v^2}} \cr
\nabla^m \phi &=&\bar{e}^m_{~a}\cosh{\sqrt{m^2 \phi^2}}\cos{\sqrt{4v^2}}\left(\frac{{\cal D}^a \phi}{\cosh{\sqrt{m^2 \phi^2}}}  -2 \delta^{ab} v_b   \left( \frac{\tan{\sqrt{4v^2}}}{\sqrt{4v^2}}  \right)\right) \cr
 & & 
\eea
and can be used to covariantly constrain (as $\omega_Z^\prime =\omega_Z$ is invariant, equation (\ref{oneformvariations})) the field $v_a$ equivalent to the constraint obtained from the $v_a$ field equation as well as constrain $\phi$ to be static as obtained from the $w$ field equation.

Indeed the field equations are obtained directly from the $AdS-C_D$ invariant action.  The $w$-equation of motion is obtained as
\bea
0 &=& \frac{\delta}{\delta w(x, t)}\Gamma_{AdS-C_D} \cr
 &=&  (\det{\bar{e}^m_{~a}}) \cosh^p{\sqrt{m^2 \phi^2}} \cos{\sqrt{4v^2}} \left[ 2 \left(\frac{\partial}{\partial t}\phi \right) \left( \frac{\tan{\sqrt{4v^2}}}{\sqrt{4v^2}} \right) \right].
\eea
The $v_a$-field equation yields
\bea
0 &=& \frac{\delta}{\delta v_a(x, t)}\Gamma_{AdS-C_D} \cr
 &=& 2\det{\bar{E}} \cosh^{p+1}{\sqrt{m^2 \phi^2}} \cos{\sqrt{4v^2}} \left(\frac{{\cal D}^b \phi}{\cosh{\sqrt{m^2 \phi^2}}}-2 \delta^{bd} v_d \left( \frac{\tan{\sqrt{4v^2}}}{\sqrt{4v^2}} \right)   \right) \times \cr
 & &\qquad \times \left[ -2v_b \left( \frac{\tan{\sqrt{4v^2}}}{\sqrt{4v^2}} \right)\frac{{\cal D}^a \phi}{\cosh{\sqrt{m^2 \phi^2}}}  \right. \cr
 & &\left.\qquad\qquad +\left( \frac{{\cal D}^c \phi}{\cosh{\sqrt{m^2 \phi^2}}}2v_c \left( \frac{\tan{\sqrt{4v^2}}}{\sqrt{4v^2}} \right) -\tan^2{\sqrt{4v^2}} + \frac{\tan{\sqrt{4v^2}}}{\sqrt{4v^2}}   \right) P_{Tba} (v) \right. \cr
 & &\left.\qquad\qquad\qquad\qquad +\left( \frac{{\cal D}^c \phi}{\cosh{\sqrt{m^2 \phi^2}}}2v_c \left( \frac{\tan{\sqrt{4v^2}}}{\sqrt{4v^2}} \right)+1\right) P_{Lba} (v) \right] \cr
 & &  \cr
 & &+2\left(\frac{\partial}{\partial t}\phi \right)(\det{\bar{e}^n_{~b}}) \cosh^p{\sqrt{m^2 \phi^2}} \cos{\sqrt{4v^2}} \left[ -\bar{e}^a_{~0} \left( \frac{\tan{\sqrt{4v^2}}}{\sqrt{4v^2}} \right) \right. \cr
 & &\left. \qquad\qquad\qquad\qquad\qquad\qquad\qquad\qquad + (w -v_f \bar{e}^f_{~0} ) \frac{v_a}{v^2} \left( 1-\frac{\tan{\sqrt{4v^2}}}{\sqrt{4v^2}} \right) \right]    \cr
 & & 
\eea
Finally the $\phi$-equation of motion is obtained
\bea
0 &=& \frac{\delta}{\delta \phi(x, t)}\Gamma_{AdS-C_D} \cr
 &=& \det{\bar{E}} \cosh^{p+1}{\sqrt{m^2 \phi^2}} \cos{\sqrt{4v^2}} (1+p) m^2 \phi \left( \frac{\tanh{\sqrt{m^2 \phi^2}}}{\sqrt{m^2 \phi^2}} \right) \cr
 & &-\det{\bar{E}} \cosh^{p}{\sqrt{m^2 \phi^2}} \cos{\sqrt{4v^2}} \left( 2\bar{e}^{-1a}_{~~~m} \partial^m v_b\right) \left[ \left(\frac{\tan{\sqrt{4v^2}}}{\sqrt{4v^2}} \right) P_{Tba} (v) + P_{Lba} (v) \right]  \cr
 & &-\left[\partial^m \left( \det{\bar{E}} \bar{e}^{-1a}_{~~~m}  \right)\right] \cosh^{p}{\sqrt{m^2 \phi^2}} \cos{\sqrt{4v^2}} \left[ 2v_a \left( \frac{\tan{\sqrt{4v^2}}}{\sqrt{4v^2}}\right)\right] \cr
 & &- (\det{\bar{e}^m_{~c}}) \cosh^p{\sqrt{m^2 \phi^2}} \cos{\sqrt{4v^2}} 2\left( \dot{w} -v_a \dot{\bar{e}}^a_{~0} \right) \left(\frac{\tan{\sqrt{4v^2}}}{\sqrt{4v^2}} \right) \cr
 & &  \cr
 & &+ (\det{\bar{e}^m_{~c}}) \cosh^p{\sqrt{m^2 \phi^2}} \cos{\sqrt{4v^2}} \left( 2 \dot{v}_a {\bar{e}}^a_{~0} \right) \left(\frac{\tan{\sqrt{4v^2}}}{\sqrt{4v^2}} \right) \cr
 & &-(\det{\bar{e}^m_{~c}}) \cosh^p{\sqrt{m^2 \phi^2}} \cos{\sqrt{4v^2}} \left( \frac{v_a \dot{v}_a}{v^2} \right)2\left( {w} -v_b {\bar{e}}^b_{~0} \right) \left[ 1-\left(\frac{\tan{\sqrt{4v^2}}}{\sqrt{4v^2}} \right)\right]  .\cr
 & & 
\label{phieqofmotion}
\eea

Introducing the momentum density $\Pi (x, t)$ as
\be
\Pi (x, t) \equiv \frac{\partial \det{E}}{\partial \dot{\phi}} = \det{\bar{E}} \bar{e}^{0-1}_{~0} \cosh^p{\sqrt{m^2 \phi^2}} \cos{\sqrt{4v^2}} 2 \frac{\tan{\sqrt{4v^2}}}{\sqrt{4v^2}}(w -v_a \bar{e}^a_{~0})
\label{Pimomentum}
\ee
and likewise defining the derivative of the Lagrangian with respect to the spatial derivatives of $\phi$ as $\Pi_m (x, t)$ 
\be
\Pi_m (x, t) \equiv \frac{\partial \det{E}}{\partial \partial^m{\phi}} = \det{\bar{E}}  \cosh^p{\sqrt{m^2 \phi^2}} \cos{\sqrt{4v^2}}  \frac{\tan{\sqrt{4v^2}}}{\sqrt{4v^2}}2 v_a \bar{e}^{-1a}_{~~~m}
\ee
the $\phi$-field equation is expressed as
\bea
0 &=& \frac{\delta \Gamma_{AdS-C_D}}{\delta \phi (x, t)} \cr
 &=& -\dot{\Pi} -\partial^m \Pi_m + \det{\bar{E}} 
 \cosh^p{\sqrt{m^2 \phi^2}} \cos{\sqrt{4v^2}} \left[ \frac{\sinh{\sqrt{m^2 \phi^2}}}{\sqrt{m^2 \phi^2}} m^2 \phi \right] \times \cr
 & & \times \left\{p+1 + p \left[ \frac{{\cal D}^a \phi}{\cosh{\sqrt{m^2 \phi^2}}} 2 v_a \frac{\tan{\sqrt{4 v^2}}}{\sqrt{4 v^2}} +\frac{\bar{e}^{0-1}_{~0} \dot{\phi}}{\cosh{\sqrt{m^2 \phi^2}}} 2(w -v_a \bar{e}^a_{~0} ) \frac{\tan{\sqrt{4 v^2}}}{\sqrt{4 v^2}}  \right]    \right\} .\cr
 & & 
\eea

Applying the $w$-field equation implies that $\dot{\phi}=0$.  As expected there is no causal connection so the field has a static spatial distribution.  Applying this to the $v_m$-field equation yields the \lq\lq inverse Higgs mechanism" \cite{Ivanov:1975zq} for the (spatial) components of the $SO(p)$ vector field $v_a$
\be
\frac{{\cal D}^a \phi}{\cosh{\sqrt{m^2 \phi^2}}} = 2 \delta^{ab} v_b \left( \frac{\tan{\sqrt{4v^2}}}{\sqrt{4v^2}} \right).
\label{invhiggsmech}
\ee
Since the $\phi$ covariant derivatives have the same form, equation (\ref{covderivphi}), the static nature of the spatial distribution of the $\phi$ field and the inverse Higgs constraint for the spatial vector field $v_a$ could equivalently be covariantly imposed on the fields by $\omega_Z =0$.  Finally applying the first two field equations to the $\phi$-field equation the momentum density time dependence is obtained
\bea
\dot{\Pi} + \partial^m \Pi_m &=& \frac{\det{\bar{E}} 
 \cosh^{(p+1)}{\sqrt{m^2 \phi^2}}}{\sqrt{\cosh^2{\sqrt{m^2 \phi^2}}+{\cal D}^a \phi {\cal D}^a \phi}}
  \left[ m^2 \phi \frac{\sinh{\sqrt{m^2 \phi^2}}}{\sqrt{m^2 \phi^2}} \right] \times \cr
 & & \qquad\qquad \times
\left\{p+1 + p \left[ \frac{{\cal D}^b \phi {\cal D}^b \phi}{\cosh^2{\sqrt{m^2 \phi^2}}} \right]    \right\}   ,
\eea
with the spatial momentum $\Pi_m$ becoming
\be
\Pi_m = \frac{\det{\bar{E}} \cosh^{(p+1)}{\sqrt{m^2 \phi^2}}}{\sqrt{\cosh^2{\sqrt{m^2 \phi^2}}+{\cal D}^b \phi {\cal D}^b \phi}} \bar{e}^{-1a}_{~~ ~m} {\cal D}^a \phi  .
\ee

The Noether current for the broken translation symmetry with parameter $\zeta$ is more complex in the background $AdS-C_d$ case and no longer simply produces the $\phi$ field equation as in the Carroll space case but involves time variation of a related composite operator.  Noether's theorem has the same form as equation (\ref{NoethersThm}) (with ${\cal L}_{AdS-C_D}$ replacing ${\cal L}_C$) with the broken space translation variations given in the Appendix A.  In addition from these variations the induced general coordinate transformation matrix $G= \partial  (t^\prime, x^\prime)/\partial (t , x)$ is obtained and hence the $\det{G}^{-1}$
\be
\det{G}^{-1}= 1 + \left( 1+p +t\frac{\partial}{\partial t} +x_m \partial^m \right) \left[ m^2 \zeta \phi \frac{\tanh{\sqrt{m^2 \phi^2}}}{\sqrt{m^2 \phi^2}} \frac{\sinh{\sqrt{m^2 x^2}}}{\sqrt{m^2 x^2}} \right] .
\ee
This yields the relation $\Delta {\cal L}_{AdS-C} + {\cal L}_{AdS-C} (\frac{\partial}{\partial t} \delta t +\partial^m \delta x_m ) = 0$ and Noether's theorem is obtained
\be
0= \dot{D} + \partial^m D_m ,
\label{Ncurrent-2}
\ee
where
\bea
D &=& \Pi \delta \phi +\delta t {\cal L}_{AdS-C_D} \cr
 &=& \Pi \cosh{\sqrt{m^2 x^2}}+m^2 \phi \frac{\tanh{\sqrt{m^2 \phi^2}}}{\sqrt{m^2 \phi^2}} \frac{\sinh{\sqrt{m^2 x^2}}}{\sqrt{m^2 x^2}} \left\{ \Pi x_m \partial^m \phi -t \Pi_m \partial^m \phi \right. \cr
 & &\left. \qquad\qquad\qquad\qquad\qquad + t \det{\bar{E}} \cosh^{(p+1)}{\sqrt{m^2 \phi^2}} \cos{\sqrt{4v^2}}   \right\} \cr
D_m &=& \Pi_m \delta \phi +\delta x_m {\cal L}_{AdS-C_D} \cr
 &=& \Pi_m \cosh{\sqrt{m^2 x^2}}+m^2 \phi \frac{\tanh{\sqrt{m^2 \phi^2}}}{\sqrt{m^2 \phi^2}} \frac{\sinh{\sqrt{m^2 x^2}}}{\sqrt{m^2 x^2}} 
\left\{ t \Pi_m \dot{\phi} -x_m \Pi \dot{\phi}  \right. \cr
 & &\left. \qquad + (\Pi_m x_n -\Pi_n x_m) \partial^n \phi -x_m \det{\bar{E}} \cosh^{(p+1)}{\sqrt{m^2 \phi^2}} \cos{\sqrt{4v^2}}   \right\} . \cr
 & & 
\eea
Applying the stationary constraint, $\dot{\phi} = 0$, the broken translation symmetry currents become
\bea
D &=& \Pi \left[\cosh{\sqrt{m^2 x^2}}+m^2 \phi \frac{\tanh{\sqrt{m^2 \phi^2}}}{\sqrt{m^2 \phi^2}} \frac{\sinh{\sqrt{m^2 x^2}}}{\sqrt{m^2 x^2}} x_m \partial^m \phi \right] \cr
 & &\qquad + t m^2 \phi \frac{\tanh{\sqrt{m^2 \phi^2}}}{\sqrt{m^2 \phi^2}} \frac{\sinh{\sqrt{m^2 x^2}}}{\sqrt{m^2 x^2}} \times \cr
 & &\qquad\qquad \times\left( \det{\bar{E}} \cosh^{(p+1)}{\sqrt{m^2 \phi^2}} \cos{\sqrt{4v^2}} -\Pi_m \partial^m \phi \right) \cr
D_m &=& \Pi_m \left[\cosh{\sqrt{m^2 x^2}}+m^2 \phi \frac{\tanh{\sqrt{m^2 \phi^2}}}{\sqrt{m^2 \phi^2}} \frac{\sinh{\sqrt{m^2 x^2}}}{\sqrt{m^2 x^2}} x_n \partial^n \phi \right] \cr
 & &\qquad - x_m m^2 \phi \frac{\tanh{\sqrt{m^2 \phi^2}}}{\sqrt{m^2 \phi^2}} \frac{\sinh{\sqrt{m^2 x^2}}}{\sqrt{m^2 x^2}} \times \cr
 & &\qquad\qquad \times \left( \det{\bar{E}} \cosh^{(p+1)}{\sqrt{m^2 \phi^2}} \cos{\sqrt{4v^2}} -\Pi_n \partial^n \phi \right) .\cr
 & &  
\label{componentNcurrents}
\eea
The non-linear broken translation symmetry of $AdS-C_D$ space no longer results in the $\phi$ field equation directly but involves a composite operator current.  In the Carroll spacetime limit $m^2 \rightarrow 0$, as applied to the above currents, $D\rightarrow \Pi$ and $D_m \rightarrow \Pi_m$ and the broken translation current conservation equation is just the $\phi$ field equation as the broken translation transformation is a simple $\phi$ field shift symmetry (\ref{txphitransformations}).

Similarly combining the $L$ and $K^a$ one-forms in matrix notation as was done for the $H$ and $P^a$ one-forms, equations (\ref{matrixE}) and (\ref{matrixEe2}), the covariant derivatives are given as $\nabla^M V_A$ with $V_A = (w , v_a)$ and
\be
\tilde{\omega}_{KA} = \begin{pmatrix}
\omega_L & \omega_{Ka} \cr
\end{pmatrix}  = dX_M \nabla^M V_A =
\begin{pmatrix}
dt & dx_m \cr
\end{pmatrix}  \begin{pmatrix}
\nabla^t w & \nabla^t v_a \cr
\nabla^m w & \nabla^m v_a \cr
\end{pmatrix}  .
\ee
Noting that under the non-linear transformations (\ref{Gtransformations}) of the coordinates and the transformations of the covariant derivative one-forms with equation (\ref{Lambda}) so that
\be
\tilde{\omega}^\prime_{KA} = \begin{pmatrix}
\omega_L^\prime & \omega_{Ka}^\prime \cr
\end{pmatrix}  = 
\begin{pmatrix}
\omega_L & \omega_{Kb} \cr
\end{pmatrix}  
\begin{pmatrix}
1&0 \cr
-\theta_b & R^{-1}_{ba} \cr
\end{pmatrix} = 
\tilde{\omega}_{KB} \Lambda^B_{~A} 
\ee
implying that 
\be
(\nabla^M V_A)^\prime = G^{-1M}_{~~~~N} ( \nabla^N V_B ) \Lambda^B_{~A} .
\ee
Combining the vielbeine into the matrix vielbein $E$ as in equation (\ref{matrixE}) , its variation involves $G$ and $\Lambda$ as shown in equation (\ref{ETrans}) .  Thus calculating the trace of $\nabla V$ using $E^{-1}$ 
\be
(E^{-1A}_{~~~~M} \nabla^M V_B)^\prime = \Lambda^{-1A}_{~~~~D} (E^{-1D}_{~~~~M} \nabla^M V_C ) \Lambda^C_{~B}
\ee
an invariant is obtained
\be
({\rm Tr}[E^{-1} \nabla V])^\prime = {\rm Tr}[{E^{-1} \nabla V}] .
\ee

Applying the $w$ and $v_a$ field equation constraints, $\dot{\phi} =0=\dot{v}_a$ and implicitly the equation (\ref{invhiggsmech}), the remaining $\phi$ field equation (\ref{phieqofmotion}) can be shown to be given by the trace of the covariant derivative of $w$ and $v_a$.  The inverse vielbein with the $\dot{\phi} =0$ constraint explicit so that $e^0_{~a}=0$ is found to be 
\be
E^{-1A}_{~~~~M} = \begin{pmatrix}
e^{0-1}_{~0}  & 0 \cr
-e^{-1a}_{~~~~n}e^n_{~0} e^{0-1}_{~0} & e^{-1a}_{~~~m} \cr
\end{pmatrix} ,
\ee
where $e^{-10}_{~~~0}=e^{0-1}_{~0}$ with $e^0_{~0} = \bar{e}^0_{~0} \cosh{\sqrt{m^2 \phi^2}}$ and the inverse submatrix $e^{-1a}_{~~~m}$ is found to be
\be
e^{-1a}_{~~~m} = \frac{1}{\cosh{\sqrt{m^2 \phi^2}}}\left[ P_{Tab} (v) +\cos{\sqrt{4v^2}}P_{Lab} (v) \right] \bar{e}^{-1b}_{~~~m} ,
\ee
both with the field constraints having been used.  Multiplying these matrices together yields the vector field covariant derivative trace
\be
{\rm Tr}[E^{-1} \nabla V] = e^{0-1}_{~0} \nabla^t w + e^{-1a}_{~~~m} \nabla^m v_a ,
\ee
where again the constraints have been used, particularly $\dot{v}_a =0$.  Applying the covariant derivatives which are recalled in equations (\ref{covderivwv}) with again the constraints applied finally yields the trace 
\bea
{\rm Tr}[E^{-1} \nabla V] &=& -\frac{1}{2} m^2 \phi (1+p) \cos{\sqrt{4 v^2}} \frac{\tanh{\sqrt{m^2 \phi^2}}}{\sqrt{m^2 \phi^2}}  \cr
 & & +\frac{1}{\cosh{\sqrt{m^2 \phi^2}}} \left(  \bar{e}^{0-1}_{~0} (\dot{w} -p \bar{\omega}_{0a}^{~t} v_a)\frac{\sin{\sqrt{4v^2}}}{\sqrt{4v^2}} \right. \cr
 & &\left. \quad + \bar{e}^{-1b}_{~~~n} \partial^n v_a  \left[ \frac{\sin{\sqrt{4v^2}}}{\sqrt{4v^2}} P_{Tab} (v) +\cos{\sqrt{4v^2}}P_{Lab} (v) \right]\right) . \cr
 & & 
\eea
Using the background one-forms, equation (\ref{backgrdvielbeine}), and background spin connections, equation (\ref{backgspinconnections}), it is found that 
\be
\partial^m \left[ \det{\bar{e}} \bar{e}^0_{~0} \bar{e}^{-1a}_{~~~m} \right] = \det{\bar{e}} \left [ \dot{\bar{e}}^{a}_{~0} - p \bar{\omega}_{0b}^{~t} \delta^{ba}  \right] ,
\ee
with $\bar{e}^a_{~0} = \bar{e}^{-1a}_{~~~~r} \bar{e}^r_{~0}$ so that $\dot{\bar{e}}^a_{~0} = \bar{e}^{-1a}_{~~~~r} \dot{\bar{e}}^r_{~0}$.  Applying this and the field constraints to the $\phi$ field equation (\ref{phieqofmotion}) and exploiting the background spin connections $\bar{\omega}_{ab}^{~n}$ and $\bar{\omega}_{0a}^{~t}$, it is found that (\ref{phieqofmotion}) (with $\dot{\phi}=0=\dot{v}_a$) can be written as
\be
0 = \frac{\delta \Gamma_{AdS-C_D}}{\delta \phi} = -2\det{\bar{E}} \cosh^{(1+p)}{\sqrt{m^2 \phi^2}}~~ {\rm Tr} [E^{-1} \nabla V] .
\label{EinvdelV}
\ee
\pagebreak
\setcounter{newapp}{3}
\setcounter{equation}{0}

\section{The Nambu-Goto Carrollian Vielbein and the Dual Vector Action}

Returning to the unconstrained fields and combining the vielbeine in matrix form $E^M_{~A}$, it and the Lagrangian, $\det{E}$, can be factorized into a product of the background $AdS_d$ world volume vielbein, $\bar{E}^M_{~A}$, times the Nambu-Goto Carrollian vielbein, $N^B_{~A}$, as such
\be
E^M_{~A} = \bar{E}^M_{~B} N^B_{~A} ,
\label{EbarENambu}
\ee
with equation (\ref{matrixEe2})
\be
E^M_{~A} = \begin{pmatrix}
e^0_{~0} & e^0_{~a} \cr
e^m_{~0} & e^m_{~a} \cr
\end{pmatrix}
\ee
and background $AdS_d$ vielbein
\be
\bar{E}^M_{~A} = \begin{pmatrix}
\bar{e}^0_{~0} & \bar{e}^0_{~a}=0 \cr
\bar{e}^m_{~0} & \bar{e}^m_{~a} \cr
\end{pmatrix}
\ee
and correspondingly the inverse background world volume vielbein
\be
\bar{E}^{-1A}_{~~~~M} = \begin{pmatrix}
\bar{e}^{0-1}_{~0} & 0 \cr
-\bar{e}^{-1a}_{~~~~r}\bar{e}^r_{~0}\bar{e}^{0-1}_{~0} & \bar{e}^{-1a}_{~~~m} \cr
\end{pmatrix}  .
\ee
This yields the Nambu-Goto Carrollian vielbein $N^B_{~A} = \bar{E}^{-1B}_{~~~M} E^M_{~A}$
\be
N^B_{~A}= \begin{pmatrix}
N^0_{~0} & N^{0}_{~a} \cr
N^b_{~0} & N^b_{~a} \cr
\end{pmatrix}
\ee
with
\bea
N^0_{~0} &=& \bar{e}^{0-1}_{~0} e^0_{~0} \cr
 &=& e^A \left[ 1+ 2w e^{-A} {\cal D}^t \phi \frac{\sin{2v}}{2v}  \right] \cr
N^b_{~0} &=& \bar{e}^{-1b}_{~~~n} \left[ e^n_{~0} - \bar{e}^n_{~0} \bar{e}^{0-1}_{~0} e^0_{~0} \right] \cr
 &=& 2w \frac{\sin{2v}}{2v} \left[ {\cal D}^b \phi - \bar{e}^b_{~0} {\cal D}^t \phi \right] + e^A w v_b \left( \frac{\cos{2v} -1}{v^2} \right)  \cr
N^0_{~a} &=& \bar{e}^{0-1}_{~0} e^0_{~a} \cr
 &=& 2{\cal D}^t \phi v_a \frac{\sin{2v}}{2v}  \cr
N^b_{~a} &=& \bar{e}^{-1b}_{~~~m} \left[ e^m_{~a} - \bar{e}^m_{~0} \bar{e}^{0-1}_{~0} e^0_{~a} \right] \cr
 &=& 2v_a \frac{\sin{2v}}{2v} \left[ {\cal D}^b \phi - \bar{e}^b_{~0} {\cal D}^t \phi  \right]+ e^A \left[ P_{Tba} (v) + \cos{2v} P_{Lba} (v) \right] , \cr
 & & 
\eea
where the notation uses the simplifing expressions
\bea
e^A &=& e^{\ln{\cosh{\sqrt{m^2 \phi^2}}}}=\cosh{\sqrt{m^2 \phi^2}} = \cosh{m\phi}  \cr
{\cal D}^t \phi &=& \bar{e}^{0-1}_{~0} \frac{\partial}{\partial t}{\phi} = \bar{e}^{0-1}_{~0} \dot{\phi} \cr
{\cal D}^a \phi &=& \bar{e}^{-1a}_{~~~m} \frac{\partial}{\partial x_m}{\phi} =\bar{e}^{-1a}_{~~~m} \partial^m {\phi} \cr
x^2 &=& x_m \delta^{mn} x_n = x_m x_m \cr
v &=& \sqrt{v^2} = \sqrt{v_a v_a} ,
\label{notation}
\eea
where recall the shorthand expressions such as $v_a = \delta^{ab} v_b$ and likewise $x_m = \delta^{mn} x_n$ are used and recall the background component vielbeine
\bea
\bar{e}^{-1a}_{~~~m} \bar{e}^m_{~0} &=& \frac{x_a t}{x^2} (1-\bar{e}^0_{~0} ) = \bar{e}^a_{~0}  \cr
\bar{e}^{-1a}_{~~~m} &=& \bar{e}^{0-1}_{~0} P_{Tam} (x) +P_{Lam} (x) \cr
\bar{e}^{0}_{~0} &=& \frac{\sinh{\sqrt{m^2 x^2}}}{\sqrt{m^2 x^2}} .
\eea

Hence the $AdS-C_D$ action is given by
\be
\Gamma_{AdS-C_D} = \int dt d^p x \det{E} = \int dt d^p x \det{\bar{E}} \det{N} 
\ee
where $\det{\bar{E}} = \bar{e}^0_{~0} \det{\bar{e}^m_{~a}}$ and
\be
\det{N} =\left( \det{N^c_{~d}} \right) \left[ N^0_{~0} - N^0_{~a} N^{-1a}_{~~~~b} N^b_{~0} \right] . 
\ee
Letting $N^a_{~b} = u_av_b + e^A \left( P_{Tab} (v) + \cos{2v} P_{Lab} (v)  \right)$ the inverse  submatrix $N^{-1a}_{~~~~b}$ is
\be
N^{-1a}_{~~~~b} = \alpha u_a v_b + e^{-A} \left[ P_{Tab} (v) +\beta P_{Lab} (v) \right]
\ee
where
\bea
\alpha &=& \frac{- e^{-A}}{[u_a v_a + e^A \cos{2v}]}  \cr
\beta &=& \frac{(e^A + u_b v_b)}{[u_a v_a + e^A \cos{2v}]}  .
\eea
For the case at hand 
\be
u_a = 2 \left( {\cal D}^a \phi - \bar{e}^a_{~0} {\cal D}^t \phi \right) \frac{\sin{2v}}{2v} 
\ee
and so $u_a v_a = 2 \left( {\cal D}^a \phi - \bar{e}^a_{~0} {\cal D}^t \phi \right)v_a \frac{\sin{2v}}{2v}$.
The submatrix determinant, $\det{N^a_{~b}}$, is also determined to be
\be
\det{N^a_{~b}} = e^{pA} \cos{2v} \left[ 1+ 2 e^{-A} \left( {\cal D}^a \phi - \bar{e}^a_{~0} {\cal D}^t \phi \right)v_a \frac{\tan{2v}}{2v} \right]  .
\ee
Putting all these expressions together yields the Nambu-Goto Carrollian determinant
\bea
\det{N} &=& \cosh^p{m\phi} \left[ \cosh{m\phi} \cos{2v} + 2 \left( {\cal D}^a \phi v_a \frac{\sin{2v}}{2v} \right) \right.  \cr
 & &\left. \qquad\qquad\qquad\qquad +2 \left( w -v_a \bar{e}^{-1a}_{~~~m}\bar{e}^m_{~0} \right){\cal D}^t \phi \frac{\sin{2v}}{2v}  \right].
\label{N-G-C}
\eea
Thus with $\det{\bar{E}} = \bar{e}^0_{~0} \det{\bar{e}^m_{~a}} = (\bar{e}^0_{~0})^p$, the $\det{E}= \det{\bar{E}}\det{N}$ Lagrangian is in agreement with the Lagrangian obtained in equation (\ref{ELagrangian}).  The invariant action is obtained
\bea
\Gamma_{AdS-C_D} &=& \int dt d^p x \det{E} = \int dt d^p x \det{\bar{E}} \det{N} \cr
 &=& \int dt d^p x \bar{e}^0_{~0} (\det{\bar{e}}) e^{pA} \left[ e^A \cos{2v} + 2 \left( {\cal D}^a \phi v_a \frac{\sin{2v}}{2v} \right) \right.  \cr
 & &\left. \qquad\qquad\qquad\qquad +2 \left( w -v_a \bar{e}^{-1a}_{~~~m} \bar{e}^m_{~0} \right){\cal D}^t \phi \frac{\sin{2v}}{2v}  \right]. 
\label{componentAdSCaction}
\eea

As in the $AdS$ case \cite{Clark:2005ht} a vector field dual formulation of the $AdS-C_D$ action can be obtained by introducing a vector field $F_M$, with $M=0,1,2,\ldots , p$
\be
F_M \equiv 2 \det{\bar{E}} \frac{\sin{2v}}{2v} V_A \bar{E}^{-1A}_{~~~M} ,
\label{FM}
\ee
with space-time component fields
\be
V_A = (w,~v_a) 
\ee
and 
\be
F_M = (f, ~f_m) .
\ee
It is useful to introduce the singular tangent space metric
\be
h^{AB} = \begin{pmatrix}
0& 0 \cr
0 & \delta^{ab} \cr
\end{pmatrix}
\ee
as well as the singular background $AdS-C_d$ metric
\bea
\bar{G}^{MN} &\equiv& \bar{E}^M_{~A} h^{AB} \bar{E}^N_{~B}  \cr
 &=& \begin{pmatrix}
0 & 0\cr
0 & \bar{E}^m_{~a} \delta^{ab} \bar{E}^n_{~b} \cr
\end{pmatrix} \cr
 &\equiv& \begin{pmatrix}
0 & 0 \cr
0 & \bar{g}^{mn} \cr
\end{pmatrix} .
\eea
Note that $\det{\bar{G}} =0$ and hence has no inverse.  Using the singular metric it is found that
\be
F_M \bar{G}^{MN} F_N = f_m \bar{g}^{mn} f_n  .
\ee
In terms of component fields it is obtained that
\be
F_M = \begin{pmatrix}
f\cr
f_m \cr
\end{pmatrix}
= 2\det{\bar{E}} \frac{\sin{2v}}{2v} \begin{pmatrix}
(w - v_a \bar{e}^{-1a}_{~~~~r} \bar{e}^r_{~0} ) \bar{e}^{0-1}_{~0} \cr
v_a \bar{e}^{-1a}_{~~~~m}  \cr
\end{pmatrix},
\ee
Inverting equation (\ref{FM})
\be
V_A = \left( \frac{1}{2 \det{\bar{E}} \frac{\sin{2v}}{2v}} \right) F_M \bar{E}^M_{~A}  ,
\ee
so that the component fields are related by 
\be
V_A = \begin{pmatrix}
w\cr
v_a \cr
\end{pmatrix}
= \frac{1}{2\det{\bar{E}} \frac{\sin{2v}}{2v}} \begin{pmatrix}
f \bar{e}^0_{~0} + f_m \bar{e}^m_{~0} \cr
f_m \bar{e}^m_{~a} \cr
\end{pmatrix}.
\ee

Hence the useful expressions are found
\bea
v^2 &=& v_a \delta^{ab} v_b = V_A h^{AB} V_B = \frac{1}{\left( 2 \det{\bar{E}} \frac{\sin{2v}}{2v} \right)^2} F_M \bar{G}^{MN} F_N   \cr
 &=& \frac{1}{\left( 2 \det{\bar{E}} \frac{\sin{2v}}{2v} \right)^2} f_m \bar{g}^{mn} f_n  ,
\eea
while 
\be
\det{\bar{E}} \cos{2v} = \sqrt{\left(\det{\bar{E}} \right)^2 - F_M\bar{G}^{MN} F_N} 
\ee
and
\bea
V_A \bar{E}^{-1A}_{~~~M} \partial^M \phi &=& \left( w - v_a \bar{e}^{-1a}_{~~~m}  \bar{e}^m_{~0} \right) {\cal D}^t \phi + v_a {\cal D}^a \phi \cr
 &\equiv& V_A \hat{{\cal D}}^A \phi  ,
\eea
where $\hat{{\cal D}}^A \phi = \bar{E}^{-1A}_{~~~M} \partial^M \phi$ with $\partial^M = (\partial^t, \partial^m)$ and the partial covariant derivatives ${\cal D}^t \phi = \bar{e}^{0-1}_{~0} \dot{\phi}$ and ${\cal D}^a \phi = \bar{e}^{-1a}_{~~~m} \partial^m {\phi}$.  

The determinant of the Nambu-Goto Carrollian vielbein, equation (\ref{N-G-C}), can be written in terms of $\phi$ and $F_M$ as
\be
\det{\bar{E}} \det{N} = e^{pA} \left( e^A \sqrt{(\det{\bar{E}})^2 - F_M \bar{G}^{MN} F_N} + F_M \partial^M \phi \right).
\ee
Following reference \cite{Clark:2005ht}, introduce the function $h(\phi)$ so that
\be
\partial^M \phi e^{pA(\phi)} \equiv \partial^M h(\phi) = \frac{dh}{d\phi} \partial^M \phi 
\ee
and hence $\frac{dh(\phi)}{d\phi} = e^{pA(\phi)}$, the $AdS-C_D$ action now has the form, after integration by parts,
\be
\Gamma_{AdS-C_D}= \int dt d^p x   \left[ e^{(p+1)A(\phi)} \sqrt{(\det{\bar{E}})^2 - F_M \bar{G}^{MN} F_N} - h(\phi) \partial^M F_M \right].
\label{actionbyparts}
\ee

The scalar field $\phi$ equation of motion follows ($d=(1+p)$)
\be
\frac{\delta \Gamma_{AdS-C_D}}{\delta \phi}=0=\frac{dh}{d\phi} \left[ d m^2 \phi \frac{\sinh{m\phi}}{m\phi}\sqrt{(\det{\bar{E}})^2 - F_M \bar{G}^{MN} F_N} -\partial^M F_M  \right] .
\ee
The $\phi$ equation of motion can be enforced by introducing the Lagrange multiplier field $L$ yielding the action
\bea
\Gamma_{AdS-C_D} &=& \int dt d^p x \left[ \sqrt{(\det{\bar{E}})^2 - F_M \bar{G}^{MN} F_N} \left( T(\phi) + L dm^2 \phi \frac{\sinh{m\phi}}{m\phi} \right)\right.  \cr
 & &\left. \qquad\qquad\qquad -L \partial^M F_M  \right] ,
\label{ActionwithL}
\eea
with
\be
T(\phi) = e^{(1+p)A(\phi)} - h(\phi) dm^2 \phi \frac{\sinh{m\phi}}{m\phi} .
\label{TF}
\ee
Thus we see that the previous $\phi$ equation of motion, now coming from the $L$ field equation, results in the $L$ dependent terms cancelling and the $\phi$ field itself being expressed in terms of $F_M$.  Thus using this equation of motion
\bea
\sinh{m\phi} &=& \frac{\partial^M F_M}{dm \sqrt{(\det{\bar{E}})^2 - F_R \bar{G}^{RS} F_S}} \cr
 &=& \frac{\dot{f} +\partial^m f_m}{dm \sqrt{(\det{\bar{E}})^2 - f_r \bar{g}^{rs} f_s}}
\label{phi-F}
\eea
the $T(\phi)$ can be written in terms of $F_M$, so adopting the notation $T(\phi)= T(\phi (F^M)) \rightarrow T(F)$, the dual vector form of the action is
\bea
\Gamma_{AdS-C_D} &=& \int dt d^p x T(F) \sqrt{(\det{\bar{E}})^2 - F_M \bar{G}^{MN} F_N} \cr
 &=& \int dt d^p x T(f, f_m) \sqrt{(\det{\bar{E}})^2 - f_m \bar{g}^{mn} f_n} 
\label{dualFaction}
\eea
along with equations (\ref{phi-F}) and (\ref{TF}) to determine $\phi= \phi(F)$ and $T(F)$.

The equivalence runs in reverse as well, introducing a Lagrange multiplier $L$ to enforce equation (\ref{phi-F}) yields the action of equation (\ref{ActionwithL}) where $T(\phi)$ is given in equation (\ref{TF}).  The fields $\phi$, $F_M$ and $L$ are independent.  The $\phi$ equation of motion allows $L$ 
to be elimiated as $\delta \Gamma_{AdS-C_D}/\delta \phi =0$ implies
\be
L= -\frac{1}{dm^2}e^{-A} \frac{dT}{d\phi}= h(\phi) .
\ee
Upon substitution into $\Gamma_{AdS-C_D}$, equation (\ref{ActionwithL}), yields equation (\ref{actionbyparts}).  The definition of $F_M$ in terms of $V_A$, equation (\ref{FM}), can be applied along with (recalling that $V^2=V_A h^{AB} V_B = v^2$)
\be
\cos{2V} =\cos{2v} = \frac{1}{(\det{\bar{E}})} \sqrt{(\det{\bar{E}})^2 - F_M \bar{G}^{MN} F_N} . 
\ee
Thus substituting this into equation (\ref{actionbyparts}) and integrating by parts results in the Nambu-Goto-Carrollian action, note equation (\ref{N-G-AdS-Action}), 
\be
\Gamma_{AdS-C_D} = \int dt d^p x  \det{\bar{E}} e^{pA(\phi)} \left[ e^{A(\phi)} \cos{2V} + \frac{\sin{2V}}{2V} 2V_A \bar{E}^{-1A}_{~~~M} \partial^M \phi \right] .
\label{appbaction}
\ee
This can be expanded in terms of component fields $w$ and $v_a$ to obtain the Nambu-Goto Carrollian action of equation (\ref{componentAdSCaction}).
\pagebreak
\setcounter{newapp}{4}
\setcounter{equation}{0}

\section{Conclusion}

A $p$-brane with codimension one was embedded in $D=d+1$-dimensional AdS-Carroll space by means of the coset method.  The vanishing speed of light, $c\to 0$, Wigner-In\"on\"u contraction of the $AdS$ space $SO(2, D-1)$ symmetry algebra was obtained and the $AdS-C_D /ISO(p)$ coset element, equation (\ref{AdSCcoset}), was formed along with the unbroken background $AdS-C_d /ISO(p)$ coset element equation (\ref{backgroundcoset}).  The non-linearly realized spontaneously broken $AdS-C_D \to AdS-C_d$ symmetry transformations were obtained in Appendix A.   The invariant brane action was found using the Maurer-Cartan one-forms.  Expanding the one-forms in terms of coordinate differentials, the vielbeine and background vielbeine were obtained.  These component vielbeine were re-assembled as a Carroll spacetime matrix vielbein, equation (\ref{backgroundmatrixE}) and (\ref{matrixE}).  The $AdS-C_D$ $SO(2, D-1)$ invariant action was shown to be given by the determinant of this matrix vielbein
\be
\Gamma_{AdS-C_D} = \int dt d^p x \det{E}   
\ee
with $\det{E}$ expressed in terms of the component fields in equation (\ref{ELagrangian}).  

The $w$, $v_a$ and $\phi$ field equations followed directly from the action.  The $w$ and $v_a$ equations of motion implied the inverse Higgs mechanism constraints yielding the static nature of the spatial shape of the $\phi$ field, $\dot{\phi} =0$, as expected from the contraction of the light cone to the time axis as $c\to 0$.  As well the auxiliary vector field $v_a$ and the spatial derivatives of $\phi$ were related, equation (\ref{invhiggsmech}).  Both of these constraints can alternatively be imposed by the invariant $\phi$-covariant derivative constraint $\omega_Z = 0$.  The canonical momentum density defined in equation (\ref{Pimomentum}) on the other hand exhibits time variation related to the shape of the brane, $\phi ( x_m)$, as well as the $AdS-C$ geometry.  Finally the $\phi$-field equation can be written in terms of the covariant derivatives of the auxiliary fields $w$ and $v_a$ as equation (\ref{EinvdelV}).

In the flat Minkowski space limit, $m^2 \to 0$, the $AdS-C_D$ results describe the $p$-brane embedded in a Carroll spacetime, $AdS-C_D \to C_D$.  These results agree with those of the more informally derived results discussed in the introduction.  In addition the broken translation symmetry Noether current in the $AdS-C$ case, equation (\ref{Ncurrent-2}), with currents equation (\ref{componentNcurrents}), go over to the Carroll space currents $z=\Pi$ , $z_m =\Pi_m$, found in Table 1.  As well the Carrollian component vielbeine can be obtained as the $m^2 \to 0$ limit of equation (\ref{vielbeine}) and correspondingly the $m^2 \to 0$ action $\Gamma_C = \int dt d^p x {\cal L}_C$ with $\det{E} \stackrel{c\to 0}{\longrightarrow} {\cal L}_C$ and action $\Gamma_C$ of equation (\ref{CMinkowski}).

The $AdS-C_D$ vielbein $E$ has the product form of the $AdS-C_d$ background vielbein $\bar{E}$ times the Nambu-Goto-Carroll vielbein $N$, $E= \bar{E} N$, as expressed in equations (\ref{EbarENambu})- (\ref{notation}).  The $p$-brane action can be re-formulated in terms of its dual vector field $F_M$ action equation (\ref{dualFaction}) with functions $\phi= \phi(F)$, equation (\ref{phi-F}), and $T(F)$, equation (\ref{TF}).  Likewise the dual action can be reformulated to yield the brane Nambu-Goto-Carrollian action equation (\ref{appbaction}) with component form equation (\ref{componentAdSCaction}).

An equivalent approach to obtain the $p$-brane action is to expose the speed of light in the already known $AdS_{(d+1)} \to AdS_d$ brane action results and take the $c\to 0$ Carrollian limit thereof.  This method was presented in Appendix B where the Maurer-Cartan one-forms and Nambu-Goto vielbein of reference \cite{Clark:2005ht} were used to obtain the $AdS$ action with the speed of light parameter, equation (\ref{N-G-AdS-Action}).  The Carrollian limit was then taken to obtain the coset method component action (\ref{componentAdSCaction}) results.

\begin{flushleft}
{\bf\large Acknowledgments}
\end{flushleft}
The work of TtV was supported in part by the NSF under grant PHY- 1102585. TtV gratefully acknowledges the hospitality of E. Bergshoeff and the Van Swinderen Institute for Particle Physics and Gravity at the University of Groningen while on sabbatical leave from Macalester College. TtV also thanks Joaquim Gomis for useful discussions.

\pagebreak

\setcounter{newapp}{1}
\setcounter{equation}{0}
\renewcommand{\theequation}{\thenewapp.\arabic{equation}}

\section*{\large Appendix A: \,  $AdS-C$ Transformations}

Using the group multiplication laws as applied to the coset $\Omega$ the non-linearly realized $AdS-C_D$ transformations are determined from
\be
g \Omega (x, t) = \Omega^\prime (x^\prime, t^\prime) h(x, t) 
\ee
where the infinitesimal $AdS-C_D$ transformations form the group elements
\be
g = e^{i\epsilon H} e^{ia_m P^m} e^{i\zeta Z} e^{i\lambda L} e^{i\kappa_m K^m} e^{\frac{i}{2}\alpha_{mn} M^{mn}} e^{i\beta_m B^m} ,
\ee
while the transformed coset element is given by
\be
\Omega^\prime (x^\prime ,t^\prime ) = e^{it^\prime H + ix_m^\prime P^m} e^{i\phi^\prime (x^\prime , t^\prime)Z} e^{iw^\prime (x^\prime , t^\prime) L + v_m^\prime (x^\prime , t^\prime) K^m}. 
\ee
The $h(x,t)$ is an element of the invariant $ISO(p)$ subgroup
\be
h(x,t) = e^{\frac{i}{2}\theta_{mn} (x,t) M^{mn}} e^{i\theta_m (x, t) B^m}
\ee
with parameters $\theta_{mn}$ and $\theta_m$ that also depend on $g$.  The transformations of the spacetime coordinates and fields are found to be non-linearly realized 
\bea
t^\prime &=& t \left[ 1+ \frac{a_m x_m}{x^2}\left( 1-\sqrt{m^2 x^2}\coth{\sqrt{m^2 x^2}} \right)\right. \cr
 & &\left. \qquad -m^2 \zeta \phi \frac{\tanh{\sqrt{m^2 \phi^2}}}{\sqrt{m^2 \phi^2}} \frac{\sinh{\sqrt{m^2 x^2}}}{\sqrt{m^2 x^2}} \right. \cr
 & &\left. \qquad\qquad -\frac{\kappa_m x_m}{x^2} 2\phi \frac{\tanh{\sqrt{m^2 \phi^2}}}{\sqrt{m^2 \phi^2}} \left( \cosh{\sqrt{m^2 x^2}}-\frac{\sqrt{m^2 x^2}}{\sinh{\sqrt{m^2 x^2}}} \right)\right] \cr
 & & + \epsilon \sqrt{m^2 x^2} \coth{\sqrt{m^2 x^2}} -\lambda 2 \phi \frac{\tanh{\sqrt{m^2 \phi^2}}}{\sqrt{m^2 \phi^2}} \frac{\sqrt{m^2 x^2}}{\sinh{\sqrt{m^2 x^2}}} - \beta_m x_m  \cr
 & & \cr
x^\prime_m &=& x_m \left[1-m^2 \zeta \phi \frac{\tanh{\sqrt{m^2 \phi^2}}}{\sqrt{m^2 \phi^2}} \frac{\sinh{\sqrt{m^2 x^2}}}{\sqrt{m^2 x^2}} \right] +\alpha_{mn} x_n \cr
 & &\qquad\qquad + \left(\sqrt{m^2 x^2} \coth{\sqrt{m^2 x^2}}P_{Tmn} (x) +P_{Lmn} (x)  \right) a_n \cr
 & & -2 \phi \frac{\tanh{\sqrt{m^2 \phi^2}}}{\sqrt{m^2 \phi^2}} \left[\frac{\sqrt{m^2 x^2}}{\sinh{\sqrt{m^2 x^2}}} P_{Tmn} (x) + \cosh{\sqrt{m^2 x^2}} P_{Lmn} (x)  \right]  \kappa_n \cr
 & & \cr
\phi^\prime (x^\prime , t^\prime ) &=& \phi (x, t) +\zeta \cosh{\sqrt{m^2 x^2}} + 2\kappa_m x_m \frac{\sinh{\sqrt{m^2 x^2}}}{\sqrt{m^2 x^2}} \cr
 & & \cr
w^\prime (x^\prime , t^\prime ) &=& w (x, t) + \epsilon m^2 x_m v_m \frac{\tanh{\sqrt{m^2 x^2}/2}}{\sqrt{m^2 x^2}}  -\beta_m v_m \cr
 & & -m^2 t \frac{\tanh{\sqrt{m^2 x^2}/2}}{\sqrt{m^2 x^2}} a_m v_m + \lambda \frac{\sqrt{4v^2}\cot{\sqrt{4v^2}}}{\cosh{\sqrt{m^2 \phi^2}}} \cr
 & &+ \lambda 2 \phi \frac{v_m x_m}{x^2} \left( \sqrt{m^2 x^2}\tanh{\sqrt{m^2 x^2}/2} \right)\left(\frac{ \tanh{\sqrt{m^2 \phi^2}}}{\sqrt{m^2 \phi^2}}  \right) \cr
 & &+ \frac{1}{2} m^2 \zeta \frac{1}{\cosh{\sqrt{m^2 \phi^2}}} \frac{\sinh{\sqrt{m^2 x^2}}}{\sqrt{m^2 x^2}} \left[ w \frac{x_m v_m}{v^2} + \sqrt{4v^2}\cot{\sqrt{4v^2}} (t -w \frac{x_m v_m}{v^2} ) \right] \cr
 & &+w \frac{1}{\cosh{\sqrt{m^2 \phi^2}}} \frac{[1 - \sqrt{4v^2}\cot{\sqrt{4v^2}}]}{v^2} v_m [P_{Tmn}(x) + \cosh{\sqrt{m^2 x^2}} P_{Lmn} (x) ] \kappa_n \cr
 & &-m^2 t \kappa_m v_m \phi \left( \frac{\tanh{\sqrt{m^2 \phi^2}}}{\sqrt{m^2 \phi^2 }} \right)  \left(  \frac{\tanh{\sqrt{m^2 x^2}}/2}{\sqrt{m^2 x^2 }/2} \right) \cr
 & &+ t \kappa_m x_m \left(\frac{\sqrt{4v^2}\cot{\sqrt{4v^2}}}{\cosh{\sqrt{m^2 \phi^2}}}\right)\left( \frac{\cosh{\sqrt{m^2 x^2}}-1}{x^2}\right)\cr
 & & \cr
v_m^\prime (x^\prime , t^\prime ) &=& v_m (x, t) + \alpha_{mn} v_n+ \frac{m^2}{2} (a_m x_n -a_n x_m ) \frac{2}{\sqrt{m^2 x^2}} \tanh{\sqrt{m^2 x^2}/2} v_n \cr
 & &+\frac{m^2}{2} \zeta \frac{1}{\cosh{\sqrt{m^2 \phi^2}}} \frac{\sinh{\sqrt{m^2 x^2}}}{\sqrt{m^2 x^2}}
\left[ \sqrt{4v^2}\cot{\sqrt{4v^2}} P_{Tmn} (v) +P_{Lmn} (v) \right] x_n  \cr
 & &-m^2 \phi \frac{\tanh{\sqrt{m^2 \phi^2}}}{\sqrt{m^2 \phi^2}} \frac{2}{\sqrt{m^2 x^2}} \tanh{\sqrt{m^2 x^2}/2} ( x_m (\kappa_n v_n) -\kappa_m (x_n v_n)) \cr
 & &+\left( \sqrt{4v^2}\cot{\sqrt{4v^2}} P_{Tmn} (v) +P_{Lmn} (v)  \right) \frac{1}{\cosh{\sqrt{m^2 \phi^2}}}\cr
 & &\qquad\qquad\qquad\qquad \left( P_{Tnr} (x) + \cosh{\sqrt{m^2 x^2}} P_{Lnr} (x) \right) \kappa_r   . 
\eea
The invariant $ISO(p)$ subgroup parameters $\theta_{mn}$ and $\theta_m$ are also obtained
\bea
\theta_{mn} &=& \alpha_{mn}+\frac{m^2}{2} (a_m x_n -a_n x_m ) \frac{2}{\sqrt{m^2 x^2}} \tanh{\sqrt{m^2 x^2}/2} \cr
 & &- m^2 \zeta \frac{1}{\cosh{\sqrt{m^2 \phi^2}}}\frac{\sinh{\sqrt{m^2 x^2}}}{\sqrt{m^2 x^2}}(x_m v_n -x_n v_m ) \frac{\tan{\sqrt{v^2}}}{\sqrt{v^2}} \cr
 & &-m^2 \phi \frac{\tanh{\sqrt{m^2 \phi^2}}}{\sqrt{m^2 \phi^2}} \frac{2}{\sqrt{m^2 x^2}} \tanh{\sqrt{m^2 x^2}/2} ( x_m \kappa_n - x_n \kappa_m ) \cr
 & &-2 \frac{\tan{\sqrt{v^2}}}{\sqrt{v^2}} \frac{1}{\cosh{\sqrt{m^2 \phi^2}}} \left[ \left(P_{Tmr} (x) +\cosh{\sqrt{m^2 x^2}} P_{Lmr} (x) 
\right)  \kappa_r v_n \right. \cr
 & &\left.\qquad\qquad - \left(P_{Tnr} (x) +\cosh{\sqrt{m^2 x^2}} P_{Lnr} (x) 
\right)  \kappa_r v_m   \right]  \cr
 & & \cr
\theta_m &=& \beta_m -\epsilon m^2 x_m \frac{\tanh{\sqrt{m^2 x^2}/2}}{\sqrt{m^2 x^2}} +m^2 t a_m \frac{\tanh{\sqrt{m^2 x^2}/2}}{\sqrt{m^2 x^2}}  \cr
 & &+ m^2 \zeta \frac{\tan{\sqrt{v^2}}}{\sqrt{v^2}} \frac{1}{\cosh{\sqrt{m^2 \phi^2}}}\frac{\sinh{\sqrt{m^2 x^2}}}{\sqrt{m^2 x^2}}(t v_m -x_m w ) \cr
 & &-\lambda m^2 x_m \phi \frac{\tanh{\sqrt{m^2 \phi^2}}}{\sqrt{m^2 \phi^2}} \frac{2}{\sqrt{m^2 x^2}} \tanh{\sqrt{m^2 x^2}/2} +2 \lambda v_m \frac{\tan{\sqrt{v^2}}}{\sqrt{v^2}} \frac{1}{\cosh{\sqrt{m^2 \phi^2}}} \cr
 & &+m^2 t \phi \frac{\tanh{\sqrt{m^2 \phi^2}}}{\sqrt{m^2 \phi^2}} \frac{2}{\sqrt{m^2 x^2}} \tanh{\sqrt{m^2 x^2}/2} \kappa_m   \cr
 & &+2t v_m (\kappa_n x_n) \frac{\tan{\sqrt{v^2}}}{\sqrt{v^2}} \frac{1}{\cosh{\sqrt{m^2 \phi^2}}} \frac{\cosh{\sqrt{m^2 x^2}}-1}{x^2}  \cr
 & &-2w \frac{\tan{\sqrt{v^2}}}{\sqrt{v^2}} \frac{1}{\cosh{\sqrt{m^2 \phi^2}}} \left( P_{Tmn} (x) + \cosh{\sqrt{m^2 x^2}} P_{Lmn} (x) \right) \kappa_n  .
\eea

The symmetry transformations for $D=d+1$ Carrollian spacetime as given in equations (\ref{txphitransformations}) and (\ref{wvtransformations}) and as well the induced local rotations and boosts
\be
\Lambda=
\begin{pmatrix}
1 & 0\cr
-\theta_n~~ & R_{nm}^{-1}  ,
\end{pmatrix}
\ee
with $R_{nm}^{-1} = \delta_{nm} -\theta_{nm}$ where the induced infinitesimal rotation has parameter $\theta_{mn}$
\be
\theta_{nm} = \alpha_{nm} -2\frac{\tan{\sqrt{v^2}}}{\sqrt{v^2}} \left( \kappa_n v_m -\kappa_m v_n  \right)
\label{thetanm}
\ee
while the unbroken induced boosts have parameter $\theta_n$
\be
\theta_n = \beta_n +2 \left( \lambda v_n -w \kappa_n  \right) \frac{\tan{\sqrt{v^2}}}{\sqrt{v^2}} 
\label{thetan}
\ee
are obtained as the $m^2 \to 0$ limit of these $AdS-C_D \to AdS-C_d$ transformations.

\pagebreak
\setcounter{newapp}{2}
\setcounter{equation}{0}
\renewcommand{\theequation}{\thenewapp.\arabic{equation}}

\section*{\large Appendix B: \, $AdS \stackrel{c \to 0}{\longrightarrow}AdS-C$}

The purpose of this appendix is to make the speed of light $c$ explicit in the $AdS_{d+1} \rightarrow AdS_d$ isometry algebra and associated coset elements in order to implement the $c\rightarrow 0$ limit directly, reproducing the action of sections 2 and 3.  Returning to the $SO(2, d)$ symmetry algebra for $AdS_{d+1}$, equation (\ref{AdSD}), where now the $SO(2,d)\rightarrow SO(2, d-1)$ isometry algebra for $AdS_{d+1} \rightarrow AdS_d$ is denoted with hatted operators so that
\bea
P^M &=&\hat{P}^M ~~{\rm for}~~ M= 0,1,2,\ldots, p \cr
P^{p+1}&=& -\hat{Z} 
\eea
and 
\bea
M^{MN}&=& \hat{M}^{MN}~~ {\rm for}~~ M,N = 0,1,2,\ldots ,p \cr
M^{p+1M} &=&\hat{K}^M ~~~~{\rm for}~~ M=0,1,2,\ldots , p ,
\eea
where now $M, N= 0,1,2,\ldots ,p$ labelling only the $AdS_d$ components while the $(p+1)^{th}$ components are separated into $\hat{Z}$ and $\hat{K}^M$.  The $SO(2,d)$ algebra becomes that used in equation (B.5) of reference \cite{Clark:2005ht}
\bea
[\hat{M}^{MN} , \hat{M}^{RS}] &=& -i\left(\eta^{MR} \hat{M}^{NS} -\eta^{MS} \hat{M}^{NR}+\eta^{NS} \hat{M}^{MR} -\eta^{NR} \hat{M}^{MS}  \right) \cr
[\hat{M}^{MN} , \hat{P}^{L}] &=& i\left( \hat{P}^{M}\eta^{NL} - \hat{P}^{N}\eta^{ML} \right) \cr
[\hat{M}^{MN} , \hat{K}^{L}] &=& i\left( \hat{K}^{M}\eta^{NL} - \hat{K}^{N}\eta^{ML} \right) \cr
[\hat{M}^{MN} , \hat{Z}] &=& 0 \qquad\qquad~\qquad;\qquad [\hat{P}^M , \hat{K}^N ]= i \eta^{MN} \hat{Z} \cr
[\hat{P}^{M} , \hat{P}^N] &=& -im^2 \hat{M}^{MN} \qquad;\qquad [\hat{P}^M , \hat{Z} ]= -i m^2 \hat{K}^M \cr
[\hat{K}^{M} , \hat{K}^N] &=& i \hat{M}^{MN} \qquad\qquad;\qquad [\hat{Z}, \hat{K}^M ]= i \hat{P}^M .
\label{SO2palgebra}
\eea

To make the speed of light explicit introduce the generators
\bea
\hat{P}^0 &=& \frac{1}{c}H \cr
\hat{P}^m &=& P^m \cr
\hat{K}^0 &=& \frac{1}{2c}L \cr
\hat{K}^m &=& \frac{1}{2}K^m \cr
\hat{Z} &=& -Z\cr
\hat{M}^{m0} &=& \frac{1}{c}B^m \cr
\hat{M}^{mn} &=& M^{mn} ,
\eea
where the spatial indices are labelled by $m, n= 1,2,\ldots , p=(d-1)$.  Hence the $SO(2,d)$ algebra of equation (\ref{SO2palgebra}) is as given in equations (\ref{algebra1})-(\ref{algebra3}) except for the four commutators involving the explicit factor of the speed of light, which are now
\bea
[B^m , B^n ] &=& -ic^2 M^{mn} \qquad\qquad ; \qquad\qquad [B^m , L ] = i c^2 K^m \cr
[B^m , H ] &=& i c^2 P^m \qquad\qquad\qquad ; \qquad\qquad  [H , L ] = -2 i c^2 Z .
\eea

Define the operators ${\cal P}^M$, ${\cal M}^{MN}$, ${\cal Z}$, and ${\cal K}^M$ with the explicit speed of light factors removed as
\bea
{\cal P}^M &=& \begin{pmatrix}
H\cr
P^m
\end{pmatrix}\cr
{\cal K}^M &=& \begin{pmatrix}
L\cr
K^m
\end{pmatrix}\cr
{\cal Z} &=& Z \cr
{\cal M}^{MN} &=& \begin{pmatrix}
0 & -B^n\cr
B^m & M^{mn}
\end{pmatrix}.
\eea
The relation to the hatted operators is given succinctly by 
\bea
{\cal P}^M &=& C^M_{~N} \hat{P}^M \cr
{\cal K}^M &=& 2C^M_{~N} \hat{K}^M \cr
{\cal Z} &=& -\hat{Z} \cr
{\cal M}^{MN} &=& C^M_{~R} \hat{M}^{RS} C^N_{~S} ,
\label{generators}
\eea
with 
\be
C^M_{~N} = \begin{pmatrix}
c & 0\cr
0 & \delta_{mn} 
\end{pmatrix}.
\ee
In terms of these operators the $SO(2,d)$ algebra of equation (\ref{SO2palgebra}) becomes
\bea
[{\cal M}^{MN} , {\cal M}^{RS}] &=& -i\left( n^{MR} {\cal M}^{NS} -n^{MS} {\cal M}^{NR}+n^{NS} {\cal M}^{MR} -n^{NR} {\cal M}^{MS}  \right) \cr
[{\cal M}^{MN} , {\cal P}^{L}] &=& i\left( {\cal P}^{M}n^{NL} - {\cal P}^{N}n^{ML} \right) \cr
[{\cal M}^{MN} , {\cal K}^{L}] &=& i\left( {\cal K}^{M}n^{NL} - {\cal K}^{N}n^{ML} \right) \cr
[{\cal M}^{MN} , {\cal Z}] &=& 0 ~~\qquad\qquad~\qquad;\qquad [{\cal P}^M , {\cal K}^N ]= -2i n^{MN} {\cal Z} \cr
[{\cal P}^{M} , {\cal P}^N] &=& -im^2 {\cal M}^{MN} ~\qquad;\qquad [{\cal P}^M , {\cal Z} ]= +\frac{i}{2} m^2 {\cal K}^M \cr
[{\cal K}^{M} , {\cal K}^N] &=& 4i {\cal M}^{MN} \qquad\qquad;\qquad [{\cal Z}, {\cal K}^M ]= -2i {\cal P}^M ,
\label{SO2palgebracexplicit}
\eea
where the metric has the form of a $(p+1)\times (p+1)$ diagonal matrix denoted $n^{MN}$
\be
n^{MN} \equiv C^M_{~R} \eta^{RS} C^N_{~S} = \begin{pmatrix}
c^2 & 0\cr
0 & -\delta_{mn}
\end{pmatrix}.
\ee

Rather than use the coset method directly with this form of the algebra, the Maurer-Cartan one-forms found using the hatted form of the algebra can be converted to one-forms with the explicit powers of $c$ exhibited and then the $c\to 0$ limit performed.  First the coset elements for the two sets of operators are identified.  Consider the coordinates
\be
\hat{x}_M \equiv (x_0 ~,~ x_m)= (ct~,~ x_m )
\ee
and
\be
X_M \equiv (t~,~ x_m)
\label{X}
\ee
that is $X_M = \hat{x}_N C^{-1N}_{~~~M}$.  Hence the coset elements
\be
e^{i\hat{x}_M \hat{P}^M} = e^{iX_M {\cal P}^M} .
\ee
Likewise let $\hat{\phi}(\hat{x}) \hat{Z} = \phi (X) {\cal Z}$ so that 
\be
e^{i\hat{\phi} \hat{Z}}= e^{i\phi {\cal Z}} ,
\ee
and $\phi = - \hat{\phi}$.  Also define the components of $\hat{v}_M$ as
\be
\hat{v}_M = (\hat{v}_0 , \hat{v}_m ) = (2cw , 2v_m )
\ee
and those of $V_M$ as
\be
V_M = (w , v_m).
\label{V}
\ee
Thus $V_M = \frac{1}{2} \hat{v}_N C^{-1N}_{~~~M}$ so that
\be
e^{i\hat{v}_M \hat{K}^M} = e^{iV_M {\cal K}^M} .
\ee
Finally equating the unbroken subgroup operators $\hat{\theta}_{MN} \hat{M}^{MN} = \Theta_{MN} {\cal M}^{MN}$ with
\be
\Theta_{RS} = C^{-1M}_{~~~R} \hat{\theta}_{MN} C^{-1N}_{~~~S}
\ee
so that
\be
\Theta_{RS} = \begin{pmatrix}
0 & -\theta_s\cr
\theta_r & \theta_{rs}
\end{pmatrix}
\ee
while 
\be
\hat{\theta}_{MN} = \begin{pmatrix}
0 & -\hat{\theta}_n\cr
\hat{\theta}_m & \hat{\theta}_{mn}
\end{pmatrix}
=\begin{pmatrix}
0 & -c~{\theta}_n\cr
c~{\theta}_m & {\theta}_{mn}
\end{pmatrix}
\ee
and the subgroup elements are equal 
\be
e^{\frac{i}{2}\hat{\theta}_{MN} \hat{M}^{MN}} = e^{\frac{i}{2}\Theta_{MN} {\cal M}^{MN}} .
\ee

These coset elements so identified, 
\be
\hat{\Omega}= e^{i\hat{x}_M \hat{P}^M} e^{i\hat{\phi} \hat{Z}}e^{i\hat{v}_M \hat{K}^M} =\Omega = e^{iX_M {\cal P}^M} e^{i\phi {\cal Z}} e^{iV_M {\cal K}^M} ,
\ee
allow their respective Maurer-Cartan one-forms to be related, recalling the one-forms $\omega=-i\Omega^{-1}d\Omega$ and $\hat{\omega} = -i \hat{\Omega}^{-1} \hat{d} \hat{\Omega}$ with $\hat{d} = d\hat{x}_M \frac{\partial}{\partial \hat{x}_M} =dt\frac{\partial}{\partial t} + dx_m \frac{\partial}{\partial x_m} =dX_M \frac{\partial}{\partial X_M}=d$, the one-forms are equal $\omega = \hat{\omega}$.  Expanding them in terms of the generators with tangent space indices $A, B =0, 1, \ldots , p$ (recall world indices $M, N = 0, 1, \ldots , p$ also)
\be
\omega = \omega_A {\cal P}^A +\omega_{\cal Z} {\cal Z} +\omega_{KA} {\cal K}^A +\frac{1}{2}\omega_{AB} {\cal M}^{AB} 
\ee
and
\be
\hat{\omega}= \hat{\omega}_A \hat{P}^A + \hat{\omega}_Z \hat{Z} + \hat{\omega}_{KA} \hat{K}^A + \frac{1}{2}\hat{\omega}_{AB} \hat{M}^{AB} ,
\ee
and utilizing equation (\ref{generators}) the Maurer-Cartan one-forms are related
\bea
\omega_A &=& \hat{\omega}_B C^{-1B}_{~~~A} \cr
\omega_{\cal Z} &=& -\hat{\omega}_Z \cr
\omega_{KA} &=& \frac{1}{2}\hat{\omega}_{KB} C^{-1B}_{~~~A} \cr
\omega_{AB} &=& \hat{\omega}_{CD} C^{-1C}_{~~~A} C^{-1D}_{~~~B} .
\eea
These yield the relations for the component one-forms and the eventual $c\to 0$ relation to the one-forms of section 2.  The explicit factors of $c$ relating the component one-forms are, with $a, b = 1, 2, \ldots , p$,
\bea
\omega_0 &=& \frac{1}{c} \hat{\omega}_0  \cr
\omega_a &=&  \hat{\omega}_a   \cr
\omega_{\cal Z} &=&  -\hat{\omega}_Z \cr
\omega_{K0} &=&  \frac{1}{2c} \hat{\omega}_{K0} \cr
\omega_{Ka} &=& \frac{1}{2} \hat{\omega}_{Ka}  \cr
\omega_{a0} &=&  \frac{1}{c} \hat{\omega}_{a0}  \cr
\omega_{ab} &=&  \hat{\omega}_{ab} .
\eea
The relation to the $AdS-C$ one-forms of section 2 is found in the $c\to 0$ limit of the above, for example $\omega_0 = \frac{1}{c} \hat{\omega}_0 \stackrel{c \to 0}{\longrightarrow} \omega_H $.

Similarly for the background one-forms for which $\hat{\bar{\Omega}} = e^{i\hat{x}_A \hat{P}^A} =\bar{\Omega} = e^{iX_A {\cal P}^A}$ and so $\bar{\omega} = \hat{\bar{\omega}}$.  Expanding in terms of the generators
\be
\bar{\omega} = \bar{\omega}_A {\cal P}^A +\frac{1}{2}\bar{\omega}_{AB} {\cal M}^{AB} 
\ee
and
\be
\hat{\bar{\omega}}= \hat{\bar{\omega}}_A \hat{P}^A + \frac{1}{2}\hat{\bar{\omega}}_{AB} \hat{M}^{AB} ,
\ee
and using the relations for the one-forms
\bea
\bar{\omega}_A &=& \hat{\bar{\omega}}_B C^{-1B}_{~~~A} \cr
\bar{\omega}_{AB} &=& \hat{\bar{\omega}}_{CD} C^{-1C}_{~~~A} C^{-1D}_{~~~B} ,
\eea
these yield the component background one-form equalities 
\bea
\bar{\omega}_0 &=& \frac{1}{c} \hat{\bar{\omega}}_0 \cr
\bar{\omega}_a &=& \hat{\bar{\omega}}_a \cr
\bar{\omega}_{a0} &=& \frac{1}{c} \hat{\bar{\omega}}_{a0} \cr
\bar{\omega}_{ab} &=& \hat{\bar{\omega}}_{ab} ,
\eea
with the $AdS-C$ background one-forms of section 2 found in the $c\to 0$ limit, for example $\bar{\omega}_0 = \frac{1}{c} \hat{\bar{\omega}}_0 \stackrel{c \to 0}{\longrightarrow}  \bar{\omega}_H$.

Applying these $c$-factor conversions to the Maurer-Cartan one-form $\hat{\omega}_A$ found in equation (2.10) of reference \cite{Clark:2005ht} for charges defined with upper indices, as is the convention here,
\be
\hat{\omega}_A = -\frac{\sinh{\sqrt{\hat{v}^2}}}{\sqrt{\hat{v}^2}} \hat{v}_A \hat{d}\hat{\phi} + \cosh{\sqrt{m^2 \hat{\phi}^2}} \left[ {P}_{TAB} (\hat{v}) +\cosh{\sqrt{\hat{v}^2}} {P}_{LAB} (\hat{v})  \right] \eta^{BC} \hat{\bar{\omega}}_C ,
\ee
with the corresponding background one-form $\hat{\bar{\omega}}_A$ of reference \cite{Clark:2005ht}
\be
\hat{\bar{\omega}}_C = \left[ \frac{\sin{\sqrt{m^2 \hat{x}^2}}}{\sqrt{m^2 \hat{x}^2}} P_{TCD} (\hat{x}) + P_{LCD} (\hat{x})  \right] \eta^{DE} d\hat{x}_E  ,
\ee
yields the resulting $\omega_A$ one-form $\omega_A = \hat{\omega}_B C^{-1B}_{~~~A}$
\be
\omega_A = \frac{\sinh{\sqrt{4V^2}}}{\sqrt{4V^2}} 2V_A d\phi + \cosh{\sqrt{m^2\phi^2}} \left[ {\cal P}_{TA}^{~~B} (V) +\cosh{\sqrt{4V^2}} {\cal P}_{LA}^{~~B} (V)  \right] \bar{\omega}_B ,
\label{MConeform2}
\ee
with the projection operators
\bea
{\cal P}_{LA}^{~~B} (V) &=& P_{LAC} (V) n^{CB} = \frac{V_A V_C n^{CB}}{V_D n^{DE} V_E}\cr
{\cal P}_{TA}^{~~B} (V) &=& \delta_{A}^{~B} - {\cal P}_{LA}^{~~B} (V) ,
\eea
and where $\hat{v}^2 = \hat{v}_A \eta^{AB} \hat{v}_B = 4 V_D n^{DE} V_E \equiv 4 V^2$.  In analogous fashion the background one-form is derived $\bar{\omega}_A = \hat{\bar{\omega}}_B C^{-1B}_{~~~A}$
\be
\bar{\omega}_A = \left[ \frac{\sin{\sqrt{m^2 {X}^2}}}{\sqrt{m^2 {X}^2}} {\cal P}_{TA}^{~~M} (X) + {\cal P}_{LA}^{~~M} (X) \right] dX_M  ,
\label{backMC}
\ee
with $\hat{x}^2 = \hat{x}_M \eta^{MN} \hat{x}_N =  X_M n^{MN} X_N \equiv X^2$.  Likewise from equation (2.10) of reference \cite{Clark:2005ht}
\be
\hat{\omega}_{Z} = \cosh{\sqrt{\hat{v}^2}}\left[ \hat{d}\hat{\phi} -  \cosh{\sqrt{m^2 \hat{\phi}^2}} \hat{\bar{\omega}}_A \eta^{AB} \hat{v}_B \frac{\tanh{\sqrt{\hat{v}^2}}}{\sqrt{\hat{v}^2}} \right] ,
\ee
from which it is found that
\be
\omega_{\cal Z} = -\hat{\omega}_Z = \cosh{\sqrt{4V^2}}\left[ d\phi +  \cosh{\sqrt{m^2\phi^2}} ~\bar{\omega}_A n^{AB} 2V_B \frac{\tanh{\sqrt{4V^2}}}{\sqrt{4V^2}} \right] .
\label{coderiv2}
\ee

The vielbein ${\cal E}_{~A}^{M}$ is defined by relating the covariant differentials $\omega_A$ to the coordinate differentials $dX_M$
\be
\omega_A \equiv dX_M {\cal E}_{~A}^{M} .
\ee
Likewise
\be
\hat{\omega}_B \equiv d\hat{x}_M \hat{E}_{~A}^{M}  
\ee
and so the vielbeine are related through the one-forms $\omega_A = \hat{\omega}_B C^{-1B}_{~~~A}$ as
\be
{\cal E}_{~A}^{M} = C^{M}_{~N} \hat{E}_{~B}^{N} C^{-1B}_{~~~A} .
\label{EhatE}
\ee
Similarly for the background one-forms
\bea
\bar{\omega}_A &=& dX_M \bar{{\cal E}}_{~A}^{M} \cr
\hat{\bar{\omega}}_A &=& d\hat{x}_M \hat{\bar{{E}}}_{~A}^{M} 
\eea
and hence the related vielbeine
\be
\bar{{\cal E}}_{~A}^{M} = C^{M}_{~N} \hat{\bar{E}}_{~B}^{N} C^{-1B}_{~~~A} .
\label{barEhatbarE}
\ee

Since the one-forms $\omega_A$ and $\bar{\omega}_A$ are already obtained the vielbeine can be read off from their forms.  From equation (\ref{backMC}) the background vielbein $\bar{{\cal E}}_{~A}^{M}$ is seen to be equal to
\be
\bar{{\cal E}}_{~A}^{M} = \frac{\sin{\sqrt{m^2 {X}^2}}}{\sqrt{m^2 {X}^2}} {\cal P}_{TA}^{~~M} (X) + {\cal P}_{LA}^{~~M} (X) .
\ee
Equation (\ref{MConeform2}) with $d\phi = dX_M \frac{\partial}{\partial X_M} \phi = dX_M \partial^M \phi$ and $\bar{\omega}_B = dX_M \bar{{\cal E}}_{~B}^{M}$ provides the vielbein ${\cal E}_{~A}^{M}$
\be
{\cal E}_{~A}^{M} = \frac{\sinh{\sqrt{4V^2}}}{\sqrt{4V^2}} 2V_A \frac{\partial}{\partial X_M}\phi + \cosh{\sqrt{m^2\phi^2}} \left[ {\cal P}_{TA}^{~~B} (V) +\cosh{\sqrt{4V^2}} {\cal P}_{LA}^{~~B} (V)  \right] \bar{{\cal E}}_B^{~M} .
\ee

The speed of light can be taken to zero to obtain the results of sections 2 and 3.  Displaying the component one-forms and vielbeine, it is found for the background one-forms and vielbeine that
\bea
\bar{\omega}_0 &=& dt \bar{{\cal E}}_{~0}^{0} + dx_m \bar{{\cal E}}_{~0}^{m} \cr
 &\stackrel{c\to 0}{\longrightarrow}& dt \left( \frac{\sinh{\sqrt{m^2 x^2}}}{\sqrt{m^2 x^2}} \right) +dx_m \left( \frac{t x_m}{x^2} \left( 1-\frac{\sinh{\sqrt{m^2 x^2}}}{\sqrt{m^2 x^2}}  \right)  \right) \cr
 &=& \bar{\omega}_H = dt \bar{e}^0_{~0}  + dx_m \bar{e}^m_{~0} \cr
 & &  \cr
\bar{\omega}_a &=& dt \bar{{\cal E}}_{~a}^{0} + dx_m \bar{{\cal E}}_{~a}^{m} \cr
 &\stackrel{c\to 0}{\longrightarrow}&  dx_m \left(\frac{\sinh{\sqrt{m^2 x^2}}}{\sqrt{m^2 x^2}} P_{Tam} (x) +P_{Lam} (x) \right) \cr
 &=&\bar{\omega}_{Pa}= dt \bar{e}^0_{~a}  + dx_m \bar{e}^m_{~a}.
\eea
Thus the vielbeine components of section 2, equations (\ref{backgrdvielbeine}) and (\ref{backgroundmatrixE}), have been obtained.  In short this $c\to 0$ limit is 
\be
\bar{{\cal E}}_{~A}^{M} \stackrel{c\to 0}{\longrightarrow} \bar{E}_{~A}^{M} .
\ee

Proceeding in a similar manner for the one-forms and vielbeine their $c\to 0$ limits are obtained as those of section 2
\bea
{\omega}_0 &=& dt {{\cal E}}_{~0}^{0} + dx_m {{\cal E}}_{~0}^{m} \cr
 &\stackrel{c\to 0}{\longrightarrow}&  dt \left[ \frac{\sin{\sqrt{4v^2}}}{\sqrt{4 v^2}} 2 w \partial^t \phi+ \cosh{\sqrt{m^2 \phi^2}} \bar{{\cal E}}_{~0}^{0} \right] \cr
 & & +dx_m \left[ \frac{\sin{\sqrt{4 v^2}}}{\sqrt{4 v^2}} 2 w \partial^m \phi + \cosh{\sqrt{m^2 \phi^2}} \bar{{\cal E}}_{~0}^{m}    \right. \cr
 & &\left. \qquad\qquad + \cosh{\sqrt{m^2 \phi^2}}\left( \cos{\sqrt{4 v^2}} -1\right) \frac{w v_b}{v^2} \bar{{\cal E}}_{~b}^{m} \right]  \cr
 &=&\omega_H = dt {E}_{~0}^{0}  + dx_m {E}_{~0}^{m} \cr
 & &  \cr
{\omega}_a &=& dt {{\cal E}}_{~a}^{0} + dx_m {{\cal E}}_{~a}^{m} \cr
 &\stackrel{c\to 0}{\longrightarrow}&  dt \left[ \frac{\sin{\sqrt{4v^2}}}{\sqrt{4 v^2}} 2 v_a \partial^t \phi \right] +dx_m \left( \frac{\sin{\sqrt{4v^2}}}{\sqrt{4 v^2}} 2 v_a \partial^m \phi   \right.  \cr
 & &\left. \qquad\qquad +\cosh{\sqrt{m^2 \phi^2}} \left[ P_{Tab} (v) + \cos{\sqrt{4v^2}} P_{Lab} (v)   \right] \bar{{\cal E}}_{~b}^{m}   \right) \cr
 &=&\omega_{Pa}= dt {E}_{~a}^{0}  + dx_m {E}_{~a}^{m}.
\eea
Thus the vielbeine components of section 2, equations (\ref{vielbeine})-(\ref{matrixEe2}), have been obtained.  In short this $c\to 0$ limit is 
\be
{{\cal E}}_{~A}^{M} \stackrel{c\to 0}{\longrightarrow} {E}_{~A}^{M} .
\ee

The $c\to 0$ limit of the $\phi$ covariant derivatives are found from the $\omega_{{\cal Z}}$ one-form equation (\ref{coderiv2})
\bea
\omega_{{\cal Z}} &\equiv& dt \tilde{\nabla}^t \phi + dx_m \tilde{\nabla}^m \phi \cr
 &\stackrel{c\to 0}{\longrightarrow}& dt \left( \dot{\phi} \cos{\sqrt{4 v^2}} \right) \cr
 & &\qquad\qquad +dx_m \cos{\sqrt{4 v^2}} \left[ \partial^m \phi - \cosh{\sqrt{m^2 \phi^2}} 2v_a \bar{e}^m_{~a}  \frac{\tan{\sqrt{4v^2}}}{\sqrt{4 v^2}} \right] \cr
 &=& \omega_Z = dt {\nabla}^t \phi + dx_m {\nabla}^m \phi ,
\eea
which agree with the $\phi$ covariant derivatives in section 2 equation (\ref{covderivphi}).  Thus the same $AdS-C$ results are obtained as in the use of the coset method.

The Nambu-Goto vielbein, $\hat{N}^B_{~A}$, is defined by factoring the background vielbein from $\hat{E}^M_{~A}$ so that
\be
\hat{E}^M_{~A} = \hat{\bar{E}}^M_{~B} \hat{N}^B_{~A}
\ee
and so
\be
\hat{N}^B_{~A} = \hat{\bar{E}}^{-1B}_{~~~~M} \hat{{E}}^M_{~A} .
\ee
Correspondingly ${\cal E}^M_{~A} = \bar{{\cal E}}^M_{~B} {\cal N}^B_{~A}$ and ${\cal N}^B_{~A} = \bar{{\cal E}}^{-1B}_{~~~~M} {\cal E}^M_{~A}$.  Exploiting the relations between $\hat{E}$, $\hat{\bar{E}}$ and ${\cal E}$, ${\bar{{\cal E}}}$, it is obtained that
\be
{\cal N}^B_{~A} = C^B_{~D} \hat{N}^D_{~C} C^{-1C}_{~~~A}  .
\ee
Thus $\det{{\cal N}} = \det{\hat{N}}$ and from equations (\ref{EhatE}) and (\ref{barEhatbarE}) $\det{{\cal E}} =\det{\hat{E}}$ as well as $\det{\bar{{\cal E}}}= \det{\hat{\bar{E}}}$.  Consequently the $AdS$ invariant action, equation (3.18) of reference \cite{Clark:2005ht}, $\Gamma_{AdS} = \int dt d^p x \det{\hat{E}}$, is written in terms of $\det{{\cal E}}$ as
\be
\det{\hat{E}} = \det{\hat{\bar{E}}} \det{\hat{N}} = \det{\bar{{\cal E}}} \det{{\cal N}}= \det{{\cal E}}.
\ee
Utilizing equation (3.20) of reference \cite{Clark:2005ht} for the $\det{\hat{N}}$
\be
\det{\hat{N}} = \cosh^d{\sqrt{m^2 \hat{\phi}^2}} \cosh{\sqrt{\hat{v}^2}} \left[ 1-\left(\hat{v}_A \frac{\tanh{\sqrt{\hat{v}^2}}}{\sqrt{\hat{v}^2}} \right) \left( \frac{\hat{\bar{E}}^{-1A}_{~~~M} \hat{\partial}^M \hat{\phi}}{\cosh{\sqrt{m^2 \hat{\phi}^2}}} \right)  \right] ,
\ee
and converting $\hat{\phi}$, $\hat{v}_A$ and $\hat{x}_M$ to $\phi$, $V_A$ and $X_M$ as well as using the relation
\be
\bar{{\cal E}}^{-1A}_{~~~~M} = C^A_{~B} \hat{\bar{E}}^{-1B}_{~~~~N} C^{-1N}_{~~~M}
\ee
in order to find that
\be
C^B_{~A} \hat{\bar{E}}^{-1A}_{~~~~M} \frac{\partial}{\partial \hat{x}_M} \hat{\phi} = -\bar{{\cal E}}^{-1B}_{~~~~M} \frac{\partial}{\partial X_M} \phi  ,
\ee
the $\det{{\cal N}}$ is found
\be
\det{{\cal N}} = \cosh^d{\sqrt{m^2 {\phi}^2}} \cosh{\sqrt{4 V^2}} \left[ 1+\left(2 V_A \frac{\tanh{\sqrt{4 V^2}}}{\sqrt{4 V^2}} \right) \left( \frac{{\bar{{\cal E}}}^{-1A}_{~~~~M} {\partial}^M \phi}{\cosh{\sqrt{m^2 {\phi}^2}}} \right)  \right] .
\ee

Thus the $AdS_{d+1} \to AdS_d$ brane embedded action $\Gamma_{AdS}$ is obtained (note equation (3.20) of reference \cite{Clark:2005ht})
\bea
\Gamma_{AdS} &=& \int d^d X \det{\bar{{\cal E}}} \det{{\cal N}}  \cr
 &=& \int d^d X \det{\bar{{\cal E}}} \cosh^d{\sqrt{m^2 \phi^2}} \cosh{\sqrt{4V^2}} \left[ 1+ \left( 2V_A \frac{\tanh{\sqrt{4V^2}}}{\sqrt{4V^2}}  \right)  \right. \cr
 & &\left. \qquad\qquad\qquad\qquad\left(\frac{1}{\cosh{\sqrt{m^2 \phi^2}}} \bar{{\cal E}}^{-1A}_{~~~~M} \frac{\partial}{\partial X_M} \phi  \right) \right]  
\label{N-G-AdS-Action}
\eea
in which the explicit factors of $c$ are in the background vielbein $\bar{{\cal E}}^M_{~A}$ and its inverse $\bar{{\cal E}}^{-1A}_{~~~~M}$ (note the form of equation (\ref{appbaction}) in which the $c\to 0$ limit is already taken).  Further, taking the $c \to 0$ limit, the explicit component fields can be exhibited from equations (\ref{X})and (\ref{V}) to obtain equation (\ref{N-G-C}) of section 3 for $\det{{\cal N}}\stackrel{c\to 0}{\longrightarrow} \det{N}$ and likewise $\det{\bar{{\cal E}}}\stackrel{c\to 0}{\longrightarrow}\det{\bar{E}}= \bar{e}^0_{~0} \det{\bar{e}^m_{~a}}$.  Thus $\Gamma_{AdS} \stackrel{c\to 0}{\longrightarrow} \Gamma_{AdS-C_D}$ and equation (\ref{componentAdSCaction}) is obtained.

\newpage
\end{document}